\documentclass[aps,prl,superscriptaddress,showpacs,twocolumn]{revtex4-1}

\usepackage{graphicx}
\usepackage{amsmath}
\usepackage{dsfont}
\usepackage{xspace}
\usepackage{color}
\usepackage[usenames,dvipsnames]{xcolor}
\usepackage[colorlinks=true,linkcolor=blue,urlcolor=blue,citecolor=blue]{hyperref}

\newcommand{\ie}{i.e.\@\xspace}
\newcommand{\eg}{e.g.\@\xspace}

\newcommand{\ve}[1]{{\bf #1}}

\newcommand{\eqw}[1]{(\ref{#1})}
\newcommand{\eq}[1]{Eq.\thinspace{}(\ref{#1})}

\newcommand{\fig}[1]{Fig.\thinspace{}\ref{#1}}

\newcommand{\fc}[1]{({#1})}
\newcommand{\figc}[2]{Fig.\thinspace{}\ref{#1}\thinspace{}\fc{#2}}

\newcommand{\Fig}[1]{Figure \ref{#1}}

\renewcommand{\d}{\ket{\downarrow}}
\renewcommand{\u}{\ket{\uparrow}}

\def\ket#1{\mathinner{|{#1}\rangle}}

\begin{document}

\title{Probing real-space and time resolved correlation functions with many-body Ramsey interferometry}

\author{Michael Knap}
\email[]{knap@physics.harvard.edu}
\affiliation{Department of Physics, Harvard University, Cambridge MA 02138, USA}
\affiliation{ITAMP, Harvard-Smithsonian Center for Astrophysics, Cambridge, MA 02138, USA}

\author{Adrian Kantian}
\affiliation{DPMC-MaNEP, University of Geneva, 24 Quai Ernest-Ansermet CH-1211 Geneva, Switzerland}

\author{Thierry Giamarchi}
\affiliation{DPMC-MaNEP, University of Geneva, 24 Quai Ernest-Ansermet CH-1211 Geneva, Switzerland}

\author{Immanuel Bloch}
\affiliation{Max-Planck-Institut f{\"u}r Quantenoptik, Hans-Kopfermann-Str. 1, 85748 Garching, Germany}
\affiliation{Fakult{\"a}t f{\"u}r Physik, Ludwig-Maximilians-Universit{\"a}t M{\"u}nchen, 80799 M{\"u}nchen, Germany}

\author{Mikhail D. Lukin}
\affiliation{Department of Physics, Harvard University, Cambridge MA 02138, USA}

\author{Eugene Demler}
\affiliation{Department of Physics, Harvard University, Cambridge MA 02138, USA}

\date{\today}

\begin{abstract}

We propose to use Ramsey interferometry and single-site addressability, available
in synthetic matter such as cold atoms or trapped ions, to measure real-space and time resolved 
spin correlation functions. These correlation functions directly 
probe the excitations of the system, which makes it possible to characterize the underlying many-body states. Moreover they contain 
valuable information about phase transitions where they exhibit scale 
invariance. We also discuss experimental imperfections and show that a spin-echo protocol can 
be used to cancel slow fluctuations in the  magnetic field.
We explicitly consider examples of the two-dimensional, antiferromagnetic Heisenberg model 
and the one-dimensional, long-range transverse field Ising model to illustrate the technique.

\end{abstract}

\pacs{
47.70.Nd, 
05.30.-d, 
75.10.Jm, 
67.85.-d, 
37.10.Ty  
}

\maketitle

In condensed matter systems there exists a common framework for understanding such 
diverse probes as neutron and X-ray scattering, electron energy loss spectroscopy, 
optical conductivity, scanning tunneling microscopy, and angle resolved 
photoemission. All of these techniques can be understood  
in terms of dynamical response functions, which are Fourier transformations of retarded 
Green's functions~\cite{fetter.walecka}
\begin{equation}
G_{\text{ret}}^{AB,\mp} (t) := -\frac{i \theta(t)}{Z} \sum_n e^{-\beta E_n} \langle n | B(t) A(0) \mp A(0) B(t) | n \rangle \;.
\label{eq:gf}
\end{equation}
Here, the summation goes over all many-body eigenstates $| n\rangle$, $\beta=1/k_BT$, the partition function
$Z= \sum_n e^{-\beta E_n}$, operators are given in the Heisenberg representation  $A(t) = e^{i{\hat H} t}
A e^{ - i {\hat H} t}$ ($\hbar$ is set to one in this manuscript), signs $-(+)$ correspond to commutator(anticommutator) Green's functions, and 
$\theta(t)$ is the Heaviside function. Correlation functions provide a direct probe 
of many-body excitations and their weight, describe many-body states, and give particularly important information about quantum 
phase transitions, where they exhibit characteristic scaling forms~\cite{sachdev_quantum_2011}.

In the last few years the experimental realization of many-body systems with ultracold atoms~\cite{bloch_many-body_2008},
polar molecules~\cite{lahaye_physics_2009}, and ion chains~\cite{blatt_quantum_2012} has opened new directions for exploring quantum dynamics.
However, most dynamical studies of such ``synthetic matter'' correspond
to quench or ramp experiments: The initial state is prepared, then it undergoes 
some nontrivial evolution $|\Psi(t)\rangle = T_t e^{ -i \int_0^t dt' {\hat H}(t')} | \Psi(0)\rangle$ 
and some observable $A$ is measured $\langle A(t) \rangle = \langle \Psi (t) | A| \Psi(t) \rangle $. 
These experiments provide an exciting new direction for exploring many-body dynamics, 
but they do not give direct information about excitations of many-body systems as 
contained in dynamical response functions. 
Notable exceptions are phase or amplitude shaking of the optical lattice 
(see, \eg, \cite{kollath_spectroscopy_2006,tokuno_spectroscopy_2011,endres_higgs_2012} and 
references therein) and radio frequency spectroscopy~\cite{stewart_using_2008},
which can be understood as measuring the single particle spectral function (i.e. the imaginary 
part of the corresponding response function). However, these techniques can not be extended 
to measuring other types of correlation functions, such as spin correlation functions 
in magnetic states as realized in optical lattices or ion chains and are often carried out in a regime
far beyond linear response, which would be required to relate the measurement
to theory within Kubo formalism~\cite{fetter.walecka}.
\begin{figure}
\begin{center}
 \includegraphics[width=0.48\textwidth]{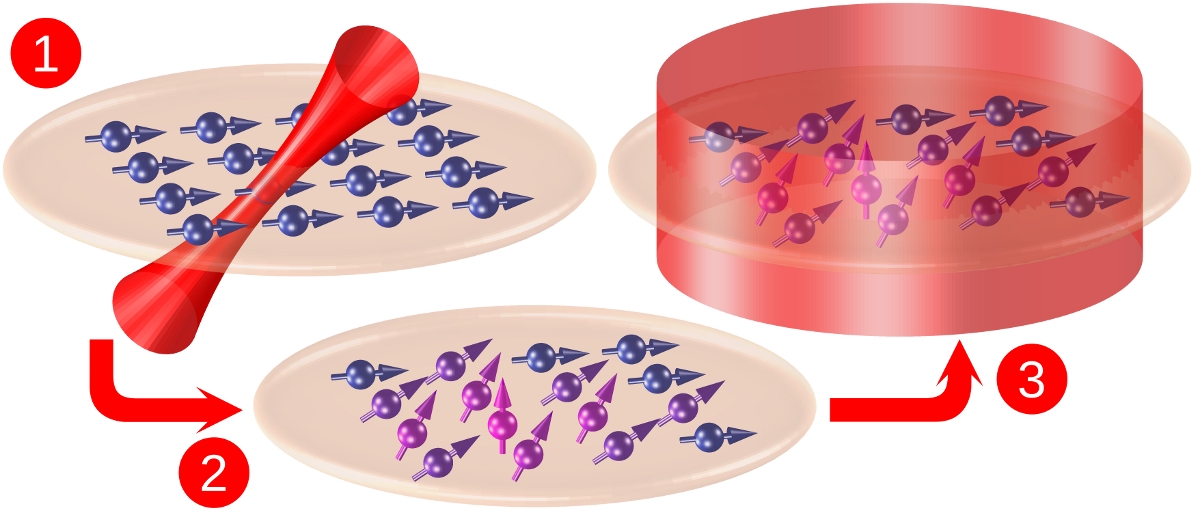}
\end{center}
\caption{\label{fig:setting} (Color online) Many-body Ramsey interferometry
consists of the following steps: (1) A spin system prepared in its ground 
state is locally excited by $\pi/2$ rotation, (2) the system evolves in time, 
(3) a global $\pi/2$ rotation is applied, followed by the measurement of the 
spin state. This protocol provides the dynamic many-body Green's function. 
}
\end{figure}

In this paper, we demonstrate that a combination of Ramsey interference experiments 
and single site addressability available in ultracold atoms and ion chains can be 
used to measure \textit{real-space and time resolved spin correlation} functions; see \fig{fig:setting}
for an illustration of the protocol. This is in contrast to established condensed matter 
probes, which generally measure response functions in frequency and wave vector domain.
In principle, the two quantities are connected by Fourier transform, but the limited bandwidth 
of experiments renders a reliable mapping difficult in practice. We further discuss 
experimental limitations such as slow magnetic field fluctuations and show that global 
spin echo can be used to cancel these fluctuations. 

\textbf{Many-body Ramsey interference.---}We consider a spin-$1/2$ system and introduce 
Pauli matrices $\sigma^a_j$ for every site $j$ with $a\in \lbrace x,y,z \rbrace$.  At this point
we do not make any assumptions on the specific form of the spin
Hamiltonian. Examples will be given below. The internal states 
$\d_z$ and $\u_z$ of a single site $j$ can be controlled by Rabi pulses which are 
of the general form~\cite{nielsen_quantum_2000,haffner_quantum_2008}
\begin{equation}
 R_j(\theta, \phi) =  \hat{\mathds{1}} \cos \frac{\theta}{2} + i(\sigma_j^x\cos \phi-\sigma_j^y\sin \phi) \sin \frac{\theta}{2}\,,
 \label{eq:rot}
\end{equation}
where $\theta = \Omega \tau$ with the Rabi frequency $\Omega$ and the pulse duration $\tau$, 
and $\phi$ the phase of the laser field. For the many-body Ramsey interference we consider 
spin rotations with $\theta=\pi/2$ but $\phi$ arbitrary. 

The many-body Ramsey protocol consists of four steps, see \fig{fig:setting} for the first three of them: (1) perform a local $\pi/2$ rotation
$R_i^1 := R_i(\pi/2, \phi_1)$ on site $i$, (2) evolve the system in time for a duration $t$, 
(3) perform a global (or local) $\pi/2$ spin rotation $R^2:=\prod_j R_j(\pi/2,\phi_2)$, and (4) measure
$\sigma^z$ on site $j$. 
The final measurement is destructive but can be carried out in parallel on all sites. 

The result of this procedure, after repetition over many experimental runs, 
corresponds to the expectation value
\begin{align}
 \label{eq:Mexp}
&M_{ij}( \phi_1,  \phi_2, t) =  \\	& \sum_n \frac{e^{-\beta E_n}}{Z}
\langle n | R^\dagger_{i}(\phi_1) e^{i {\hat H} t} R^\dagger(\phi_2)
\sigma_{j}^{z} R(\phi_2) e^{- i {\hat H} t} R_{i}(\phi_1) | n \rangle \nonumber \;.
\end{align}
With some algebra we obtain~\cite{supp}
\begin{align}
M_{ij}( &\phi_1,  \phi_2, t) = \frac{1}{2} \big( \cos \phi_1 \sin \phi_2 G_{ij}^{xx,-}
+  \cos \phi_1 \cos \phi_2 G_{ij}^{xy,-} \nonumber \\ &-  \sin \phi_1 \sin \phi_2 G_{ij}^{yx,-}
-  \sin \phi_1 \cos \phi_2 G_{ij}^{yy,-} \big)
\nonumber\\
&+ \text{terms with odd number of } \sigma^{x,y}  \text{ operators}\;,
\label{Mfull}
\end{align}
where $G_{ij}^{ab,-}$ is the retarded, commutator Green's function defined in \eq{eq:gf}
with $A=\sigma^a_i$ and $B=\sigma^b_j$.

In many physically relevant models, terms with odd number of $\sigma$ operators
vanish by symmetry or at least can be removed by an appropriate choice of the 
phases $\phi_1$ and $\phi_2$ of the laser fields. We show below when using these properties that in 
cases of both the Heisenberg model, \eq{eq:heis}, 
and the long-range, transverse field Ising model, \eq{eq:ising}, our Ramsey interference sequence measures 
a combination of retarded correlation functions
\begin{allowdisplaybreaks}
\begin{align}
M_{ij}( \phi_1,  \phi_2, t)  = \frac{1}{4}  \big\lbrace &\sin (\phi_1{+}\phi_2) ( G_{ij}^{xx,-}  {-}  G_{ij}^{yy,-} ) \nonumber\\-
&\sin (\phi_1{-}\phi_2) (  G_{ij}^{xx,-}  {+}  G_{ij}^{yy,-} )  \nonumber\\
+ &\cos (\phi_1{+}\phi_2) (   G_{ij}^{xy,-}{+}  G_{ij}^{yx,-} )\nonumber\\ 
+&\cos (\phi_1{-}\phi_2) (  G_{ij}^{xy,-} {-}  G_{ij}^{yx,-}  ) \big\rbrace \;.
\label{eq:M}
\end{align}
\end{allowdisplaybreaks}

Alternatively to the many-body Ramsey protocol, a spin-shelving technique can be used 
to measure dynamic spin correlations along the quantization direction, \ie, the operators
$A$ and $B$ in \eqw{eq:gf} are $\sigma_i^z$ and $\sigma_j^z$, respectively~\cite{supp}. In the supplementary 
material~\cite{supp} we also derive a useful relation between Green's functions and Loschmidt 
echo, and discuss that it can be used to characterize diffusive and localized many-body phases.
\begin{figure}
\begin{center}
 \includegraphics[width=0.49\textwidth]{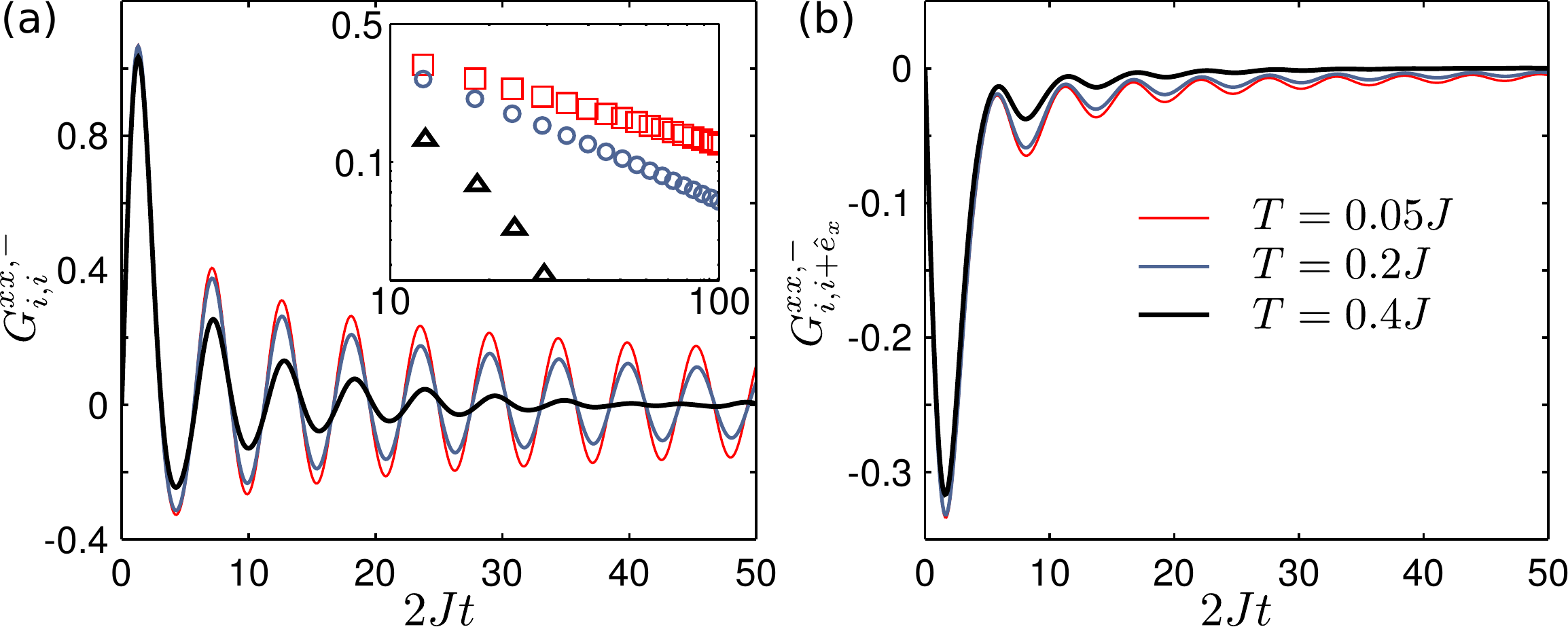}
\end{center}
\caption{\label{fig:hGF} (Color online) Real-space and time resolved 
Green's function $G^{xx,-}_{ij}$ of the two-dimensional, isotropic Heisenberg model, which can 
be measured with many-body Ramsey interferometry, shown for different
temperatures $T$. The antiferromagnetic correlations manifest
themselves in the opposite phase of on-site \fc{a} and nearest-neighbor \fc{b}
correlations. The inset in \fc{a} shows the decay of the peaks in
$G^{xx,-}_{ii}$ on a double logarithmic scale. See main text for details.
}
\end{figure}

\textbf{Heisenberg model.---}The anisotropic Heisenberg model of the XXZ type,
can be realized both with two component mixtures~\cite{rey_preparation_2007,
trotzky_time-resolved_2008,nascimbene_experimental_2012,fukuhara_quantum_2013,greif_short-range_2013,mag} 
and with polar molecules~\cite{gorshkov_tunable_2011,gorshkov_quantum_2011,yan_realizing_2013} in optical lattices
\begin{eqnarray}
{\hat H}_{\rm Heis} = \sum_{ i<j } J_{ij}^\perp ( \sigma^x_i \sigma^x_j +  \sigma^y_i \sigma^y_j )
+ J_{ij}^z \sigma_i^z \sigma_j^z \;.
\label{eq:heis}
\end{eqnarray}
For two component Bose mixtures, interactions can be mediated through the superexchange mechanism and 
$J_{ij}^\perp$, $J_{ij}^z$ are functions of the inter- and intra-species scattering 
lengths which are nonzero for $i,j$ nearest neighbors. When realizing the
Heisenberg model with polar molecules, $J_{ij}^\perp$, $J_{ij}^z$ are long-ranged and 
anisotropic in space. Hamiltonian \eqw{eq:heis}
is introduced for arbitrary dimension and the site index $i$ is understood as
a collective index. We assume that the system is prepared
in equilibrium at finite temperature, \ie, it has a density matrix given by
$\rho= Z^{-1}e^{-\beta {\hat H_{\rm Heis}}}$.

Hamiltonian \eqw{eq:heis} has the global symmetry $\sigma^x \rightarrow - \sigma^x$, 
$\sigma^y \rightarrow - \sigma^y$, and $\sigma^z \rightarrow \sigma^z$, from which it
is obvious that expectation values with an odd number of $\sigma^{x,y}$ vanish. 
In addition, Hamiltonian \eqw{eq:heis} has a U(1) symmetry of spin rotations around the $z$ axis.
This symmetry requires that 
\begin{align*}
G_{ij}^{xx} - G_{ij}^{yy} = 0 \quad \text{   and } \quad
G_{ij}^{xy} + G_{ij}^{yx}=0 \;.
\end{align*}
Hence, the many-body Ramsey protocol \eqw{eq:M} measures
\begin{align}
M_{ij}( \phi_1,  \phi_2, t) = -\frac{1}{4} \big \lbrace &\sin (\phi_1 - \phi_2) ( G_{ij}^{xx}  +  G_{ij}^{yy} ) \nonumber	 \\
  - & \cos (\phi_1 - \phi_2) ( G_{ij}^{xy}  -  G_{ij}^{yx} ) \big \rbrace \;.
\label{MshortHeisenberg}
\end{align}
The choice of the phases $\phi_1$ and $\phi_2$ of the laser fields, determines which combination
of Green's functions is obtained.  

In case the two spin states are not encoded in magnetic field insensitive states, 
one may also need to take into account fluctuating magnetic fields for a realistic 
measurement scenario. Such a contribution
is described by a Zeeman term $\hat H_\text{Z}= h_z \sum_i \sigma^z_i$.
A spin-echo sequence, however, which augments the Ramsey protocol with a global $\pi$ rotation $R^\pi$
after half of the time evolution, removes slow fluctuations in the Zeeman field
\begin{align}
 &R(\phi_2) e^{-i(\hat H_{\rm Heis}+\hat H_\text{Z}) \frac{t}{2}} R^\pi e^{-i(\hat H_{\rm Heis}+\hat H_\text{Z}) \frac{t}{2}} R_i(\phi_1) \nonumber \\
 &\to \tilde R(\phi_2) e^{-i\hat H_{\rm Heis} {t}} R_i(\phi_1) \;,
 \label{eq:seShort}
\end{align}
where $\tilde R(\phi_2) = i R(\phi_2) (\cos \phi_\pi \prod_l \sigma_l^x - \sin \phi_\pi \prod_l \sigma_l^y)$ 
and $\phi_\pi$ the phase of the laser field in the course of the $\pi$ rotation. We show in~\cite{supp} that this 
transformation still allows one to measure dynamic correlation functions.

\Fig{fig:hGF} shows the time-resolved, local \fc{a} and nearest-neighbor \fc{b} 
Green's function of the antiferromagnetic Heisenberg model 
($J_{ij}^\perp = J_{ij}^z =: J$ for $i,j \in$ nearest neighbors and 
$J_{ij}^\perp = J_{ij}^z=0$ otherwise) for different 
temperatures. We obtain the results using a large-$N$ expansion in 
Schwinger-Boson representation~\cite{arovas_functional_1988,auerbach_spin_1988,auerbach_interacting_1994},
which has been demonstrated to give reasonable results for the two-dimensional spin-$1/2$
Heisenberg antiferromagnet~\cite{manousakis_spin-_1991}.
The local and nearest-neighbor, 
dynamic Green's functions show clear signatures of antiferromagnetic order, 
since their oscillations are out of phase. When lowering the temperature, the 
emergence of quantum coherence manifests through the increase in the
amplitude of the oscillations. Further the decay of the oscillations, inset in \fc{a},
follows at low temperatures a power-law over several decades in time, which indicates the 
approach to criticality. The power-law, however, is cut off by the finite 
correlation time $\log \tau \sim J/T$. Dynamic correlations at the antiferromagnetic 
ordering wave-vector $\vec \pi:=(\pi,\pi)$ are a precursor of  long-range 
order~\cite{supp} which in two dimensions emerges at zero temperature. These
correlations can be obtained  from the spatial ones by summing up contributions 
of one sublattice with positive sign and of the other with negative sign. 

\begin{figure}
\begin{center}
 \includegraphics[width=0.44\textwidth]{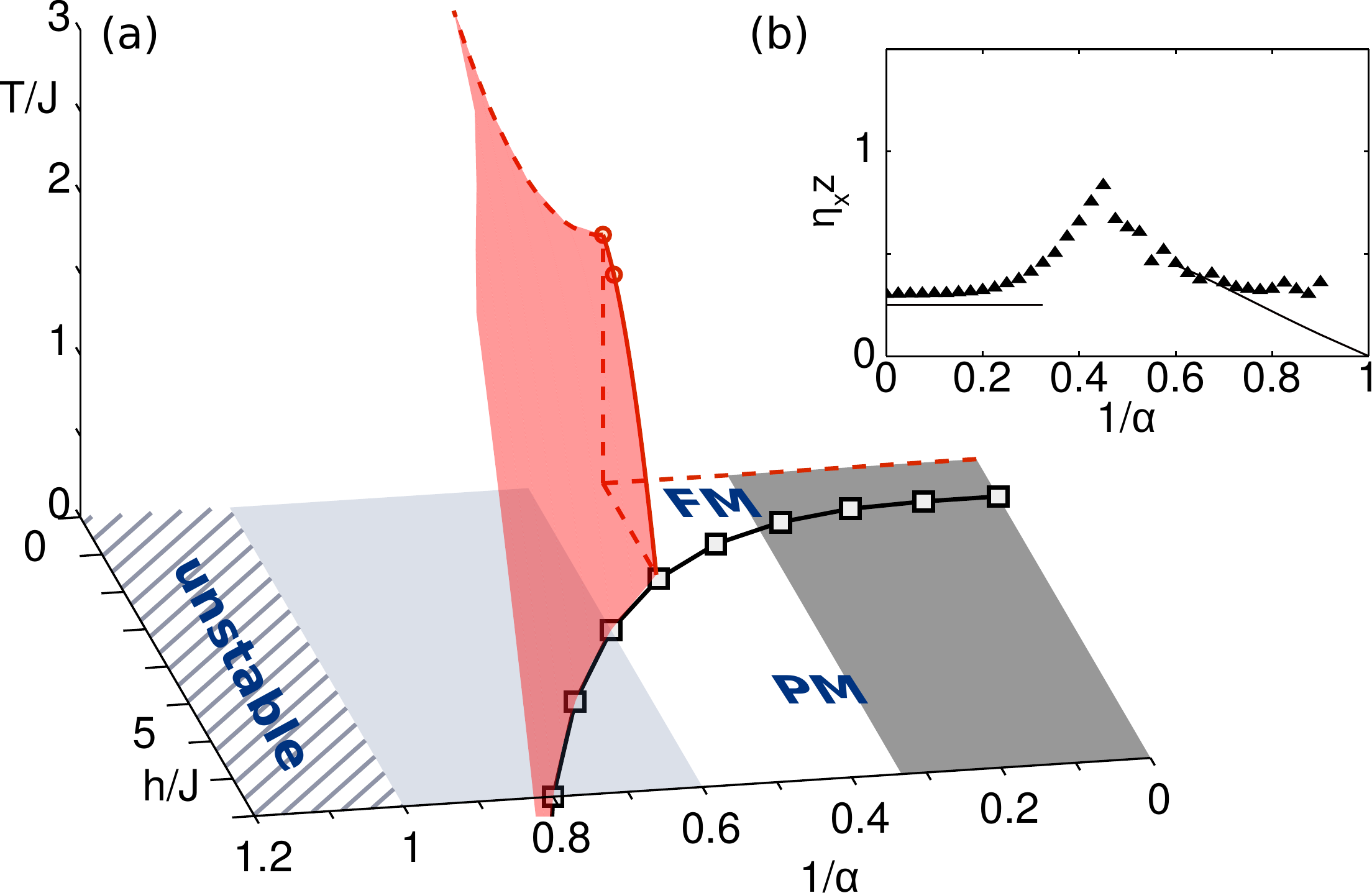}
\end{center}
\caption{\label{fig:pd} (Color online) Phase diagram \fc{a} of the one-dimensional, long-range, transverse field Ising model \eqw{eq:ising} in the 
transverse field $h$, interaction exponent $\alpha$, and temperature $T$ space. For $\alpha<1$, hatched region, the system is 
thermodynamically unstable. The solid, black line indicates the quantum critical line, which
separates the ferromagnetic (FM) and paramagnetic (PM) phase. For $\alpha>3$, dark gray region, the phase transition is 
of the same universality class as the short-range Ising transition, for $\alpha<5/3$, 
light gray region, mean-field analysis is exact~\cite{supp}. At $\alpha=2$ and $h=0$, dashed lines, the phase transition is of the 
Berezinskii-Kosterlitz-Thouless type, which also extends
to finite transverse field $h$~\cite{dutta_phase_2001}. Symbols which indicate the finite temperature 
transition correspond to $h=0$, $T=1.5262(5)J$~\cite{luijten_criticality_2001} and
$h=J/2$, $T=1.42(1)J$~\cite{sandvik_stochastic_2003}. 
\fc{b} Critical exponent $\eta_x z$ of dynamic correlations $G^{xx,-}_{L/2,L/2}(t)$ obtained along 
the critical line from the scaling of finite size systems, which are
realizable in current experiments, symbols, and exact results in the thermodynamic limit, lines.
}
\end{figure}

\textbf{Long-range, transverse field Ising model.---}Systems of trapped ions are capable of simulating canonical quantum spin models,
where two internal states of the ions serve as effective spin states and the 
interaction between spins is mediated by collective vibrations~\cite{porras_effective_2004,deng_effective_2005}.
Among the quantum spin models that can be simulated with trapped ions 
is the long-range, transverse field Ising model
\begin{equation}
 \hat H_{\text{Ising}} = - \sum_{i<j} J_{ij} \sigma_i^x \sigma_j^x - h \sum_i \sigma_i^y
 \label{eq:ising}
\end{equation}
where the spin-spin interactions fall off approximately as a power 
law $J_{ij}=J/|i-j|^\alpha$ with exponent $\alpha$, and $h$ is the 
strength of the transverse field. In trapped ion systems power-law 
interactions can be engineered 
with an exponent $\alpha$ that is highly tunable~\cite{porras_effective_2004}. 
The upper 
limit of $\alpha$ is given by the decay of dipolar interactions $0<\alpha<3$,
however, the shorter-ranged interactions are, the slower are the overall time scales, 
which in turn is challenging for experiments.

Experimentally the long-ranged Ising model has been realized with ion chains for both 
ferromagnetic (FM) $J>0$~\cite{friedenauer_simulating_2008,islam_onset_2011} as well as antiferromagnetic 
$J<0$~\cite{kim_entanglement_2009,kim_quantum_2010,edwards_quantum_2010,lanyon_universal_2011,
britton_engineered_2012,islam_emergence_2013,richerme_2013} 
coupling. Theoretically, quantum spin systems with long-range interactions
that decay with arbitrary exponent $\alpha$ have rarely been studied 
in the literature and so far static properties~\cite{dutta_phase_2001,
laflorencie_critical_2005,sandvik_ground_2010,koffel_entanglement_2012} and 
quantum quenches~\cite{wall_out_2012,hauke_spread_2013,schachenmayer_entanglement_2013} have been explored. 
This is why we discuss the one-dimensional, ferromagnetic ($J>0$), long-range, 
transverse-field Ising model in greater detail and focus in particular on 
dynamical correlation functions and on the quantum phase transition (QPT) 
from the ferromagnetic (FM) to the paramagnetic (PM) phase, whose 
universality is described by a continuous manifold of critical exponents 
that can be tuned by the decay of the interactions $\alpha$, see \figc{fig:pd}{a}
for the rich phase diagram.

The transverse field Ising model obeys the global symmetry $\sigma^x \rightarrow -\sigma^x$, $\sigma^y \rightarrow \sigma^y$ 
and $\sigma^z \rightarrow - \sigma^z$ and thus only expectation values with an odd
number of $\sigma^x$ operators vanish in \eq{eq:M}. However, when choosing the phases $\phi_1=0$ and $\phi_2=\pi/2$
it can be shown that the many-body Ramsey protocol measures~\cite{supp}
\begin{equation}
 M_{ij}(0, \pi/2, t)=\frac12 G^{xx,-}_{ij}\;.
 \label{eq:Mising}
\end{equation}

\begin{figure}
\begin{center}
 \includegraphics[width=0.48\textwidth]{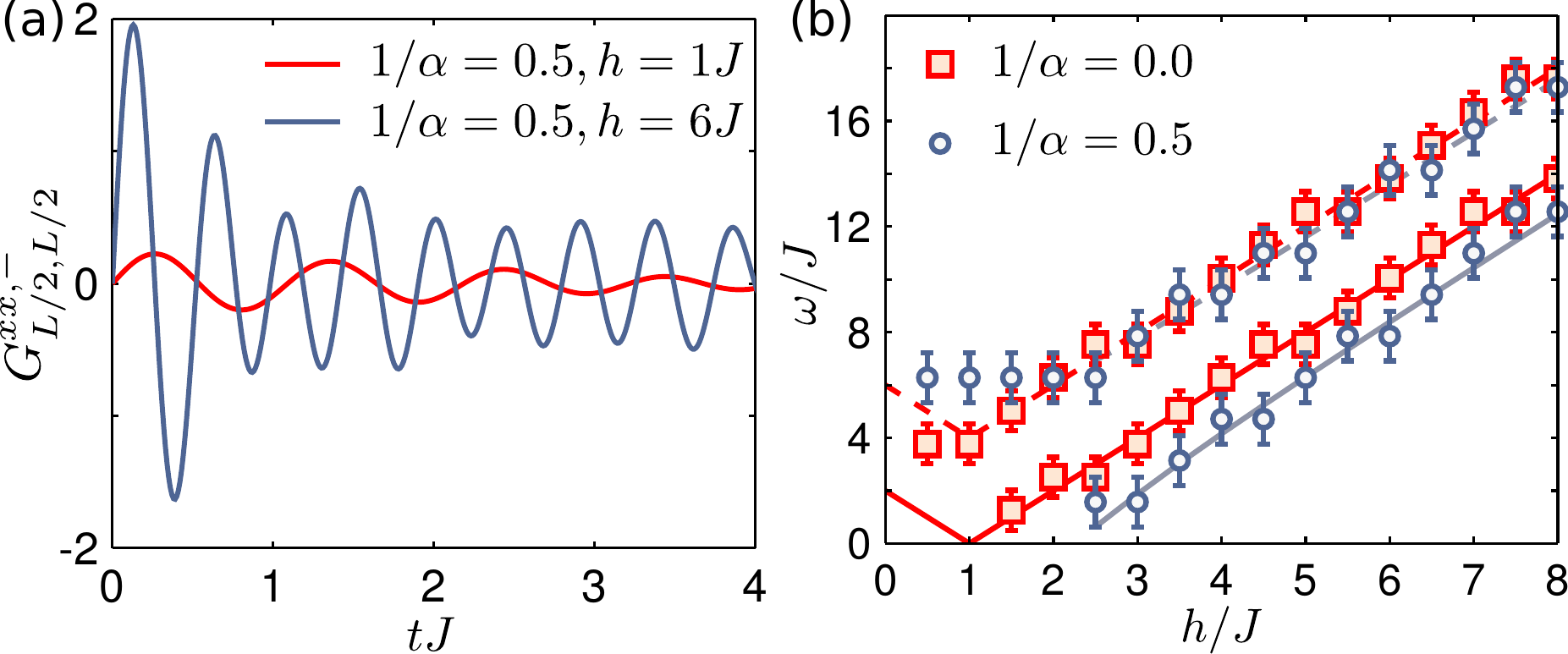}
\end{center}
\caption{\label{fig:osc} (Color online)
Dynamic Green's function $G^{xx,-}_{L/2,L/2}(t)$ \fc{a} of the long-range, transverse field Ising model \eqw{eq:ising} for interaction 
exponent $\alpha=2$ in the ferromagnetic ($h=J$) and in the paramagnetic ($h=6J$) 
phase, see legend. \fc{b} Oscillation frequencies, symbols, in 
$G^{xx,-}_{L/2,L/2}(t)$ obtained from a Fourier transform
of the time-dependent data as a function of the transverse field $h$ for two different 
values of the interaction exponent $\alpha$. Error bars indicate the resolution of the 
Fourier transform in frequency space. Solid lines illustrates the excitation gap
and dashed line the upper band edge, which define the oscillations contributing 
to the dynamic correlations. 
}
\end{figure}

We illustrate that insight into the many-body physics can be obtained by 
studying systems which are currently experimentally realizable. To this end we solve 
systems of up to 22 ions with exact diagonalization based on the Lanczos 
technique~\cite{saad_numerical_1995} and calculate their 
dynamical Green's functions. As realized in experiments we generally consider open boundary 
conditions (OBC). In \figc{fig:osc}{a} we show dynamic Green's functions 
$G^{xx,-}_{L/2,L/2}$ for the interaction exponent
$\alpha=2$ in the FM and in the PM phase. The time-resolved Green's functions
characterize the many-body states: In the FM phase ($h$ smaller than the critical 
field $h_c$ that determines the QPT) the response in the direction
of the ferromagnet is small, which manifests in $G^{xx,-}_{L/2,L/2}$ through small amplitude 
oscillations whose envelope decays very slowly, whereas in the PM phase ($h>h_c$) the response
is large, which in $G^{xx,-}_{L/2,L/2}$ manifests in oscillations that initially have a large amplitude but decay quickly in time.

The oscillations in the dynamic Green's functions contain information about the
excitations in the system. In particular, in the PM phase oscillations
with a frequency corresponding to the gap~\cite{sachdev_quantum_2011} are expected. In addition, the spectrum is 
cut off due to the lattice, which gives rise to a second energy scale present in both the PM and the FM phase.
In \figc{fig:osc}{b} we show the frequency components extracted from the Fourier transform 
of $G^{xx,-}_{L/2,L/2}(t) $ with error bars given by the resolution in frequency space for both
short-ranged interactions ($1/\alpha = 0$, squares) and the long-ranged interactions 
($1/\alpha = 1/2$, circles).  For short-range interactions $1/\alpha=0$ the gap can 
be evaluated analytically $\Delta=2|h-J|$~\cite{sachdev_quantum_2011}, which grows linearly with the transverse 
field as indicated by the solid red (dark) line in \figc{fig:osc}{b}. The upper band edge at $\Delta+4J$
is indicated by the dashed red (dark) line. At the critical point $h_c=J$ the gap closes, however 
oscillations from the finite bandwidth are still present.
For long-ranged interactions, we extract the excitation gap 
and the bandwidth numerically. Results are shown 
by blue (light) solid and dashed lines, respectively. The upper band-edge, 
blue (light) dashed line, almost coincides with the short range system. The gap 
and the upper band-edge are in good agreement with the frequency components extracted from the 
correlation functions. 

Along the quantum critical line $h=h_c(\alpha)$, which can be determined experimentally 
by measuring for example the Binder ratio~\cite{binder_critical_1981}, the system becomes scale 
invariant and thus spatial and temporal correlations decay as power laws (see \figc{fig:pd}{b}). 
In~\cite{supp} we show in detail that a  change in the critical exponents 
should be observable in current experiments already with a medium number of ions.

\textbf{Conclusions and outlook.---}In summary, we proposed a protocol to measure real-space
and time resolved spin correlation functions using many-body Ramsey interference. We discuss the 
protocol for two relevant examples of the Heisenberg and the long-range transverse field 
Ising model, which can be experimentally realized with cold atoms, polar molecules, and trapped ions.
In this work we focused on spin-$1/2$ systems. However, the proposed protocol can be 
generalized to higher-spin systems when realizing the Rabi pulses \eqw{eq:rot} with 
the respective higher-spin operators. In order to implement the generalized spin-rotations, spin states 
should be encoded in internal atomic states with isotropic energy spacing which can be 
simultaneously addressed by Rabi pulses.

The measurement of the time dependent Green's functions provides important information on
many-body excitations and on quantum phase transitions where they exhibit specific 
scaling laws. Having such tools at hand makes it possible to explore fundamental, theoretically 
much debated many-body phenomena. In particular, we 
believe that the many-body localization transition~\cite{basko_metalinsulator_2006,
gornyi_interacting_2005,oganesyan_localization_2007} and
many-body localized phases, which are characterized by a dephasing 
time that grows exponentially with the distance between two particles in the 
sample~\cite{serbyn_universal_2013,huse_phenomenology_2013,serbyn_local_2013}, can be explored using the
ideas described in this work. 

Another question is whether the many-body Ramsey protocol can be applied to 
systems out of equilibrium. The protocol we propose is based on 
discrete symmetries of many-body eigenstates and thus holds for ensembles 
described by diagonal density matrices, while a generic system 
out of equilibrium is characterized by a density matrix which also contains off-diagonal 
elements. However, if the off-diagonal elements dephase in time,
many-body Ramsey interferometry can be applied out of equilibrium as well.
This could for example also be the case for integrable systems which after fast 
dephasing are described by a diagonal density matrix whose weights are 
determined by the generalized Gibbs ensemble~\cite{rigol_relaxation_2007,rigol_thermalization_2008}.

\textbf{Acknowledgments.---}We thank S. Gopalakrishnan, R. Islam, C. Monroe, and S. Sachdev
for useful discussions. The authors acknowledge support from Harvard-MIT CUA, 
the DARPA OLE program, AFOSR MURI on Ultracold Molecules, ARO-MURI on Atomtronics, 
the Austrian Science Fund (FWF) Project No. J 3361-N20, as well as the Swiss NSF 
under MaNEP and Division II. TG  is grateful to the Harvard Physics Department and 
to Harvard-MIT CUA for support and hospitality during the completion of this work. 
Numerical calculations have been performed on the Odyssey cluster at Harvard University Research 
Computing.


\begin{thebibliography}{10}%
\makeatletter
\providecommand \@ifxundefined [1]{%
 \ifx #1\undefined \expandafter \@firstoftwo
 \else \expandafter \@secondoftwo
\fi
}%
\providecommand \@ifnum [1]{%
 \ifnum #1\expandafter \@firstoftwo
 \else \expandafter \@secondoftwo
\fi
}%
\providecommand \enquote [1]{``#1''}%
\providecommand \bibnamefont  [1]{#1}%
\providecommand \bibfnamefont [1]{#1}%
\providecommand \citenamefont [1]{#1}%
\providecommand\href[0]{\@sanitize\@href}%
\providecommand\@href[1]{\endgroup\@@startlink{#1}\endgroup\@@href}%
\providecommand\@@href[1]{#1\@@endlink}%
\providecommand \@sanitize [0]{\begingroup\catcode`\&12\catcode`\#12\relax}%
\@ifxundefined \pdfoutput {\@firstoftwo}{%
 \@ifnum{\z@=\pdfoutput}{\@firstoftwo}{\@secondoftwo}%
}{%
 \providecommand\@@startlink[1]{\leavevmode}%
 \providecommand\@@endlink[0]{}%
}{%
 \providecommand\@@startlink[1]{%
  \leavevmode
  \pdfstartlink
   attr{/Border[0 0 1 ]/H/I/C[0 1 1]}%
   user{/Subtype/Link/A<</Type/Action/S/URI/URI(#1)>>}%
  \relax
 }%
 \providecommand\@@endlink[0]{\pdfendlink}%
}%
\providecommand \url  [0]{\begingroup\@sanitize \@url }%
\providecommand \@url [1]{\endgroup\@href {#1}{\urlprefix}}%
\providecommand \urlprefix [0]{URL }%
\providecommand \Eprint[0]{\href }%
\@ifxundefined \urlstyle {%
  \providecommand \doi [1]{doi:\discretionary{}{}{}#1}%
}{%
  \providecommand \doi [0]{doi:\discretionary{}{}{}\begingroup
  \urlstyle{rm}\Url }%
}%
\providecommand \doibase [0]{http://dx.doi.org/}%
\providecommand \Doi[1]{\href{\doibase#1}}%
\providecommand \bibAnnote [3]{%
  \BibitemShut{#1}%
  \begin{quotation}\noindent
    \textsc{Key:}\ #2\\\textsc{Annotation:}\ #3%
  \end{quotation}%
}%
\providecommand \bibAnnoteFile [2]{%
  \IfFileExists{#2}{\bibAnnote {#1} {#2} {\input{#2}}}{}%
}%
\providecommand \typeout [0]{\immediate \write \m@ne }%
\providecommand \selectlanguage [0]{\@gobble}%
\providecommand \bibinfo [0]{\@secondoftwo}%
\providecommand \bibfield [0]{\@secondoftwo}%
\providecommand \translation [1]{[#1]}%
\providecommand \BibitemOpen[0]{}%
\providecommand \bibitemStop [0]{}%
\providecommand \bibitemNoStop [0]{.\EOS\space}%
\providecommand \EOS [0]{\spacefactor3000\relax}%
\providecommand \BibitemShut [1]{\csname bibitem#1\endcsname}%
\bibitem{fetter.walecka}%
  \BibitemOpen
  \bibfield{author}{%
  \bibinfo {author} {\bibfnamefont{A.~L.}\ \bibnamefont{Fetter}}\ and\ \bibinfo
  {author} {\bibfnamefont{J.~D.}\ \bibnamefont{Walecka}},\ }%
  \emph{\bibinfo {title} {Quantum Theory of Many-Particle Systems}}\ (\bibinfo
  {publisher} {McGraw-Hill},\ \bibinfo {address} {New York},\ \bibinfo {year}
  {1971})%
  \bibAnnoteFile{NoStop}{fetter.walecka}%
\bibitem{sachdev_quantum_2011}%
  \BibitemOpen
  \bibfield{author}{%
  \bibinfo {author} {\bibfnamefont{S.}~\bibnamefont{Sachdev}},\ }%
  \emph{\bibinfo {title} {Quantum Phase Transitions}},\ \bibinfo {edition}
  {2nd}\ ed.\ (\bibinfo {publisher} {Cambridge University Press},\ \bibinfo
  {address} {Cambridge, UK},\ \bibinfo {year} {2011})%
  \bibAnnoteFile{NoStop}{sachdev_quantum_2011}%
\bibitem{bloch_many-body_2008}%
  \BibitemOpen
  \bibfield{author}{%
  \bibinfo {author} {\bibfnamefont{I.}~\bibnamefont{Bloch}}, \bibinfo {author}
  {\bibfnamefont{J.}~\bibnamefont{Dalibard}},\ and\ \bibinfo {author}
  {\bibfnamefont{W.}~\bibnamefont{Zwerger}},\ }%
  \bibfield{journal}{%
  \Doi{10.1103/RevModPhys.80.885}{\bibinfo {journal} {Rev. Mod. Phys.}}\ }%
  \textbf{\bibinfo {volume} {80}},\ \bibinfo {pages} {885} (\bibinfo {year}
  {2008})%
  \bibAnnoteFile{NoStop}{bloch_many-body_2008}%
\bibitem{lahaye_physics_2009}%
  \BibitemOpen
  \bibfield{author}{%
  \bibinfo {author} {\bibfnamefont{T.}~\bibnamefont{Lahaye}}, \bibinfo {author}
  {\bibfnamefont{C.}~\bibnamefont{Menotti}}, \bibinfo {author}
  {\bibfnamefont{L.}~\bibnamefont{Santos}}, \bibinfo {author}
  {\bibfnamefont{M.}~\bibnamefont{Lewenstein}},\ and\ \bibinfo {author}
  {\bibfnamefont{T.}~\bibnamefont{Pfau}},\ }%
  \bibfield{journal}{%
  \bibinfo {journal} {Rep. Prog. Phys.}\ }%
  \textbf{\bibinfo {volume} {72}},\ \bibinfo {pages} {126401} (\bibinfo {year}
  {2009})%
  \bibAnnoteFile{NoStop}{lahaye_physics_2009}%
\bibitem{blatt_quantum_2012}%
  \BibitemOpen
  \bibfield{author}{%
  \bibinfo {author} {\bibfnamefont{R.}~\bibnamefont{Blatt}}\ and\ \bibinfo
  {author} {\bibfnamefont{C.~F.}\ \bibnamefont{Roos}},\ }%
  \bibfield{journal}{%
  \Doi{10.1038/nphys2252}{\bibinfo {journal} {Nat. Phys.}}\ }%
  \textbf{\bibinfo {volume} {8}},\ \bibinfo {pages} {277} (\bibinfo {year}
  {2012})%
  \bibAnnoteFile{NoStop}{blatt_quantum_2012}%
\bibitem{kollath_spectroscopy_2006}%
  \BibitemOpen
  \bibfield{author}{%
  \bibinfo {author} {\bibfnamefont{C.}~\bibnamefont{Kollath}}, \bibinfo
  {author} {\bibfnamefont{A.}~\bibnamefont{Iucci}}, \bibinfo {author}
  {\bibfnamefont{T.}~\bibnamefont{Giamarchi}}, \bibinfo {author}
  {\bibfnamefont{W.}~\bibnamefont{Hofstetter}},\ and\ \bibinfo {author}
  {\bibfnamefont{U.}~\bibnamefont{Schollw{\"o}ck}},\ }%
  \bibfield{journal}{%
  \Doi{10.1103/PhysRevLett.97.050402}{\bibinfo {journal} {Phys. Rev. Lett.}}\
  }%
  \textbf{\bibinfo {volume} {97}},\ \bibinfo {pages} {050402} (\bibinfo {year}
  {2006})%
  \bibAnnoteFile{NoStop}{kollath_spectroscopy_2006}%
\bibitem{tokuno_spectroscopy_2011}%
  \BibitemOpen
  \bibfield{author}{%
  \bibinfo {author} {\bibfnamefont{A.}~\bibnamefont{Tokuno}}\ and\ \bibinfo
  {author} {\bibfnamefont{T.}~\bibnamefont{Giamarchi}},\ }%
  \bibfield{journal}{%
  \Doi{10.1103/PhysRevLett.106.205301}{\bibinfo {journal} {Phys. Rev. Lett.}}\
  }%
  \textbf{\bibinfo {volume} {106}},\ \bibinfo {pages} {205301} (\bibinfo {year}
  {2011})%
  \bibAnnoteFile{NoStop}{tokuno_spectroscopy_2011}%
\bibitem{endres_higgs_2012}%
  \BibitemOpen
  \bibfield{author}{%
  \bibinfo {author} {\bibfnamefont{M.}~\bibnamefont{Endres}}, \bibinfo {author}
  {\bibfnamefont{T.}~\bibnamefont{Fukuhara}}, \bibinfo {author}
  {\bibfnamefont{D.}~\bibnamefont{Pekker}}, \bibinfo {author}
  {\bibfnamefont{M.}~\bibnamefont{Cheneau}}, \bibinfo {author}
  {\bibfnamefont{P.}~\bibnamefont{Schau{\ss}}}, \bibinfo {author}
  {\bibfnamefont{C.}~\bibnamefont{Gross}}, \bibinfo {author}
  {\bibfnamefont{E.}~\bibnamefont{Demler}}, \bibinfo {author}
  {\bibfnamefont{S.}~\bibnamefont{Kuhr}},\ and\ \bibinfo {author}
  {\bibfnamefont{I.}~\bibnamefont{Bloch}},\ }%
  \bibfield{journal}{%
  \Doi{10.1038/nature11255}{\bibinfo {journal} {Nature}}\ }%
  \textbf{\bibinfo {volume} {487}},\ \bibinfo {pages} {454} (\bibinfo {year}
  {2012})%
  \bibAnnoteFile{NoStop}{endres_higgs_2012}%
\bibitem{stewart_using_2008}%
  \BibitemOpen
  \bibfield{author}{%
  \bibinfo {author} {\bibfnamefont{J.~T.}\ \bibnamefont{Stewart}}, \bibinfo
  {author} {\bibfnamefont{J.~P.}\ \bibnamefont{Gaebler}},\ and\ \bibinfo
  {author} {\bibfnamefont{D.~S.}\ \bibnamefont{Jin}},\ }%
  \bibfield{journal}{%
  \Doi{10.1038/nature07172}{\bibinfo {journal} {Nature}}\ }%
  \textbf{\bibinfo {volume} {454}},\ \bibinfo {pages} {744} (\bibinfo {year}
  {2008})%
  \bibAnnoteFile{NoStop}{stewart_using_2008}%
\bibitem{nielsen_quantum_2000}%
  \BibitemOpen
  \bibfield{author}{%
  \bibinfo {author} {\bibfnamefont{M.~A.}\ \bibnamefont{Nielsen}}\ and\
  \bibinfo {author} {\bibfnamefont{I.~L.}\ \bibnamefont{Chuang}},\ }%
  \emph{\bibinfo {title} {Quantum Computation and Quantum Information}}\
  (\bibinfo {publisher} {Cambridge University Press},\ \bibinfo {address}
  {Cambridge, UK},\ \bibinfo {year} {2000})%
  \bibAnnoteFile{NoStop}{nielsen_quantum_2000}%
\bibitem{haffner_quantum_2008}%
  \BibitemOpen
  \bibfield{author}{%
  \bibinfo {author} {\bibfnamefont{H.}~\bibnamefont{H{\"a}ffner}}, \bibinfo
  {author} {\bibfnamefont{C.}~\bibnamefont{Roos}},\ and\ \bibinfo {author}
  {\bibfnamefont{R.}~\bibnamefont{Blatt}},\ }%
  \bibfield{journal}{%
  \Doi{10.1016/j.physrep.2008.09.003}{\bibinfo {journal} {Phys. Rep.}}\ }%
  \textbf{\bibinfo {volume} {469}},\ \bibinfo {pages} {155} (\bibinfo {year}
  {2008})%
  \bibAnnoteFile{NoStop}{haffner_quantum_2008}%
\bibitem{supp}%
  \BibitemOpen
  \bibinfo {note} {See supplementary material}%
  \bibAnnoteFile{NoStop}{supp}%
\bibitem{rey_preparation_2007}%
  \BibitemOpen
  \bibfield{author}{%
  \bibinfo {author} {\bibfnamefont{A.~M.}\ \bibnamefont{Rey}}, \bibinfo
  {author} {\bibfnamefont{V.}~\bibnamefont{Gritsev}}, \bibinfo {author}
  {\bibfnamefont{I.}~\bibnamefont{Bloch}}, \bibinfo {author}
  {\bibfnamefont{E.}~\bibnamefont{Demler}},\ and\ \bibinfo {author}
  {\bibfnamefont{M.~D.}\ \bibnamefont{Lukin}},\ }%
  \bibfield{journal}{%
  \Doi{10.1103/PhysRevLett.99.140601}{\bibinfo {journal} {Phys. Rev. Lett.}}\
  }%
  \textbf{\bibinfo {volume} {99}},\ \bibinfo {pages} {140601} (\bibinfo {year}
  {2007})%
  \bibAnnoteFile{NoStop}{rey_preparation_2007}%
\bibitem{trotzky_time-resolved_2008}%
  \BibitemOpen
  \bibfield{author}{%
  \bibinfo {author} {\bibfnamefont{S.}~\bibnamefont{Trotzky}}, \bibinfo
  {author} {\bibfnamefont{P.}~\bibnamefont{Cheinet}}, \bibinfo {author}
  {\bibfnamefont{S.}~\bibnamefont{F{\"o}lling}}, \bibinfo {author}
  {\bibfnamefont{M.}~\bibnamefont{Feld}}, \bibinfo {author}
  {\bibfnamefont{U.}~\bibnamefont{Schnorrberger}}, \bibinfo {author}
  {\bibfnamefont{A.~M.}\ \bibnamefont{Rey}}, \bibinfo {author}
  {\bibfnamefont{A.}~\bibnamefont{Polkovnikov}}, \bibinfo {author}
  {\bibfnamefont{E.~A.}\ \bibnamefont{Demler}}, \bibinfo {author}
  {\bibfnamefont{M.~D.}\ \bibnamefont{Lukin}},\ and\ \bibinfo {author}
  {\bibfnamefont{I.}~\bibnamefont{Bloch}},\ }%
  \bibfield{journal}{%
  \Doi{10.1126/science.1150841}{\bibinfo {journal} {Science}}\ }%
  \textbf{\bibinfo {volume} {319}},\ \bibinfo {pages} {295} (\bibinfo {year}
  {2008})%
  \bibAnnoteFile{NoStop}{trotzky_time-resolved_2008}%
\bibitem{nascimbene_experimental_2012}%
  \BibitemOpen
  \bibfield{author}{%
  \bibinfo {author} {\bibfnamefont{S.}~\bibnamefont{Nascimb{\`e}ne}}, \bibinfo
  {author} {\bibfnamefont{Y.-A.}\ \bibnamefont{Chen}}, \bibinfo {author}
  {\bibfnamefont{M.}~\bibnamefont{Atala}}, \bibinfo {author}
  {\bibfnamefont{M.}~\bibnamefont{Aidelsburger}}, \bibinfo {author}
  {\bibfnamefont{S.}~\bibnamefont{Trotzky}}, \bibinfo {author}
  {\bibfnamefont{B.}~\bibnamefont{Paredes}},\ and\ \bibinfo {author}
  {\bibfnamefont{I.}~\bibnamefont{Bloch}},\ }%
  \bibfield{journal}{%
  \Doi{10.1103/PhysRevLett.108.205301}{\bibinfo {journal} {Phys. Rev. Lett.}}\
  }%
  \textbf{\bibinfo {volume} {108}},\ \bibinfo {pages} {205301} (\bibinfo {year}
  {2012})%
  \bibAnnoteFile{NoStop}{nascimbene_experimental_2012}%
\bibitem{fukuhara_quantum_2013}%
  \BibitemOpen
  \bibfield{author}{%
  \bibinfo {author} {\bibfnamefont{T.}~\bibnamefont{Fukuhara}}, \bibinfo
  {author} {\bibfnamefont{A.}~\bibnamefont{Kantian}}, \bibinfo {author}
  {\bibfnamefont{M.}~\bibnamefont{Endres}}, \bibinfo {author}
  {\bibfnamefont{M.}~\bibnamefont{Cheneau}}, \bibinfo {author}
  {\bibfnamefont{P.}~\bibnamefont{Schau{\ss}}}, \bibinfo {author}
  {\bibfnamefont{S.}~\bibnamefont{Hild}}, \bibinfo {author}
  {\bibfnamefont{D.}~\bibnamefont{Bellem}}, \bibinfo {author}
  {\bibfnamefont{U.}~\bibnamefont{Schollw{\"o}ck}}, \bibinfo {author}
  {\bibfnamefont{T.}~\bibnamefont{Giamarchi}}, \bibinfo {author}
  {\bibfnamefont{C.}~\bibnamefont{Gross}}, \bibinfo {author}
  {\bibfnamefont{I.}~\bibnamefont{Bloch}},\ and\ \bibinfo {author}
  {\bibfnamefont{S.}~\bibnamefont{Kuhr}},\ }%
  \bibfield{journal}{%
  \Doi{10.1038/nphys2561}{\bibinfo {journal} {Nat. Phys.}}\ }%
  \textbf{\bibinfo {volume} {9}},\ \bibinfo {pages} {235} (\bibinfo {year}
  {2013})%
  \bibAnnoteFile{NoStop}{fukuhara_quantum_2013}%
\bibitem{greif_short-range_2013}%
  \BibitemOpen
  \bibfield{author}{%
  \bibinfo {author} {\bibfnamefont{D.}~\bibnamefont{Greif}}, \bibinfo {author}
  {\bibfnamefont{T.}~\bibnamefont{Uehlinger}}, \bibinfo {author}
  {\bibfnamefont{G.}~\bibnamefont{Jotzu}}, \bibinfo {author}
  {\bibfnamefont{L.}~\bibnamefont{Tarruell}},\ and\ \bibinfo {author}
  {\bibfnamefont{T.}~\bibnamefont{Esslinger}},\ }%
  \bibfield{journal}{%
  \Doi{10.1126/science.1236362}{\bibinfo {journal} {Science}}\ }%
  \textbf{\bibinfo {volume} {340}},\ \bibinfo {pages} {1307} (\bibinfo {year}
  {2013})%
  \bibAnnoteFile{NoStop}{greif_short-range_2013}%
\bibitem{mag}%
  \BibitemOpen
  \bibinfo {note} {The short-range, transverse field Ising
  model~\cite{simon_quantum_2011} as well as frustrated classical
  magnetism~\cite{struck_quantum_2011} have also been explored with cold
  atoms.}%
  \bibAnnoteFile{Stop}{mag}%
\bibitem{gorshkov_tunable_2011}%
  \BibitemOpen
  \bibfield{author}{%
  \bibinfo {author} {\bibfnamefont{A.~V.}\ \bibnamefont{Gorshkov}}, \bibinfo
  {author} {\bibfnamefont{S.~R.}\ \bibnamefont{Manmana}}, \bibinfo {author}
  {\bibfnamefont{G.}~\bibnamefont{Chen}}, \bibinfo {author}
  {\bibfnamefont{J.}~\bibnamefont{Ye}}, \bibinfo {author}
  {\bibfnamefont{E.}~\bibnamefont{Demler}}, \bibinfo {author}
  {\bibfnamefont{M.~D.}\ \bibnamefont{Lukin}},\ and\ \bibinfo {author}
  {\bibfnamefont{A.~M.}\ \bibnamefont{Rey}},\ }%
  \bibfield{journal}{%
  \Doi{10.1103/PhysRevLett.107.115301}{\bibinfo {journal} {Phys. Rev. Lett.}}\
  }%
  \textbf{\bibinfo {volume} {107}},\ \bibinfo {pages} {115301} (\bibinfo {year}
  {2011})%
  \bibAnnoteFile{NoStop}{gorshkov_tunable_2011}%
\bibitem{gorshkov_quantum_2011}%
  \BibitemOpen
  \bibfield{author}{%
  \bibinfo {author} {\bibfnamefont{A.~V.}\ \bibnamefont{Gorshkov}}, \bibinfo
  {author} {\bibfnamefont{S.~R.}\ \bibnamefont{Manmana}}, \bibinfo {author}
  {\bibfnamefont{G.}~\bibnamefont{Chen}}, \bibinfo {author}
  {\bibfnamefont{E.}~\bibnamefont{Demler}}, \bibinfo {author}
  {\bibfnamefont{M.~D.}\ \bibnamefont{Lukin}},\ and\ \bibinfo {author}
  {\bibfnamefont{A.~M.}\ \bibnamefont{Rey}},\ }%
  \bibfield{journal}{%
  \Doi{10.1103/PhysRevA.84.033619}{\bibinfo {journal} {Phys. Rev. A}}\ }%
  \textbf{\bibinfo {volume} {84}},\ \bibinfo {pages} {033619} (\bibinfo {year}
  {2011})%
  \bibAnnoteFile{NoStop}{gorshkov_quantum_2011}%
\bibitem{yan_realizing_2013}%
  \BibitemOpen
  \bibfield{author}{%
  \bibinfo {author} {\bibfnamefont{B.}~\bibnamefont{Yan}}, \bibinfo {author}
  {\bibfnamefont{S.~A.}\ \bibnamefont{Moses}}, \bibinfo {author}
  {\bibfnamefont{B.}~\bibnamefont{Gadway}}, \bibinfo {author}
  {\bibfnamefont{J.~P.}\ \bibnamefont{Covey}}, \bibinfo {author}
  {\bibfnamefont{K.~R.~A.}\ \bibnamefont{Hazzard}}, \bibinfo {author}
  {\bibfnamefont{A.~M.}\ \bibnamefont{Rey}}, \bibinfo {author}
  {\bibfnamefont{D.~S.}\ \bibnamefont{Jin}},\ and\ \bibinfo {author}
  {\bibfnamefont{J.}~\bibnamefont{Ye}},\ }%
  \bibfield{journal}{%
  \Doi{10.1038/nature12483}{\bibinfo {journal} {Nature (London)}}\ }%
  \textbf{\bibinfo {volume} {501}},\ \bibinfo {pages} {521} (\bibinfo {year}
  {2013})%
  \bibAnnoteFile{NoStop}{yan_realizing_2013}%
\bibitem{arovas_functional_1988}%
  \BibitemOpen
  \bibfield{author}{%
  \bibinfo {author} {\bibfnamefont{D.~P.}\ \bibnamefont{Arovas}}\ and\ \bibinfo
  {author} {\bibfnamefont{A.}~\bibnamefont{Auerbach}},\ }%
  \bibfield{journal}{%
  \Doi{10.1103/PhysRevB.38.316}{\bibinfo {journal} {Phys. Rev. B}}\ }%
  \textbf{\bibinfo {volume} {38}},\ \bibinfo {pages} {316} (\bibinfo {year}
  {1988})%
  \bibAnnoteFile{NoStop}{arovas_functional_1988}%
\bibitem{auerbach_spin_1988}%
  \BibitemOpen
  \bibfield{author}{%
  \bibinfo {author} {\bibfnamefont{A.}~\bibnamefont{Auerbach}}\ and\ \bibinfo
  {author} {\bibfnamefont{D.~P.}\ \bibnamefont{Arovas}},\ }%
  \bibfield{journal}{%
  \Doi{10.1103/PhysRevLett.61.617}{\bibinfo {journal} {Phys. Rev. Lett.}}\ }%
  \textbf{\bibinfo {volume} {61}},\ \bibinfo {pages} {617} (\bibinfo {year}
  {1988})%
  \bibAnnoteFile{NoStop}{auerbach_spin_1988}%
\bibitem{auerbach_interacting_1994}%
  \BibitemOpen
  \bibfield{author}{%
  \bibinfo {author} {\bibfnamefont{A.}~\bibnamefont{Auerbach}},\ }%
  \emph{\bibinfo {title} {Interacting electrons and quantum magnetism}}\
  (\bibinfo {publisher} {Springer},\ \bibinfo {address} {New York},\ \bibinfo
  {year} {1994})%
  \bibAnnoteFile{NoStop}{auerbach_interacting_1994}%
\bibitem{manousakis_spin-_1991}%
  \BibitemOpen
  \bibfield{author}{%
  \bibinfo {author} {\bibfnamefont{E.}~\bibnamefont{Manousakis}},\ }%
  \bibfield{journal}{%
  \Doi{10.1103/RevModPhys.63.1}{\bibinfo {journal} {Rev. Mod. Phys.}}\ }%
  \textbf{\bibinfo {volume} {63}},\ \bibinfo {pages} {1} (\bibinfo {year}
  {1991})%
  \bibAnnoteFile{NoStop}{manousakis_spin-_1991}%
\bibitem{dutta_phase_2001}%
  \BibitemOpen
  \bibfield{author}{%
  \bibinfo {author} {\bibfnamefont{A.}~\bibnamefont{Dutta}}\ and\ \bibinfo
  {author} {\bibfnamefont{J.~K.}\ \bibnamefont{Bhattacharjee}},\ }%
  \bibfield{journal}{%
  \Doi{10.1103/PhysRevB.64.184106}{\bibinfo {journal} {Phys. Rev. B}}\ }%
  \textbf{\bibinfo {volume} {64}},\ \bibinfo {pages} {184106} (\bibinfo {year}
  {2001})%
  \bibAnnoteFile{NoStop}{dutta_phase_2001}%
\bibitem{luijten_criticality_2001}%
  \BibitemOpen
  \bibfield{author}{%
  \bibinfo {author} {\bibfnamefont{E.}~\bibnamefont{Luijten}}\ and\ \bibinfo
  {author} {\bibfnamefont{H.}~\bibnamefont{Me{\ss}ingfeld}},\ }%
  \bibfield{journal}{%
  \Doi{10.1103/PhysRevLett.86.5305}{\bibinfo {journal} {Phys. Rev. Lett.}}\ }%
  \textbf{\bibinfo {volume} {86}},\ \bibinfo {pages} {5305} (\bibinfo {year}
  {2001})%
  \bibAnnoteFile{NoStop}{luijten_criticality_2001}%
\bibitem{sandvik_stochastic_2003}%
  \BibitemOpen
  \bibfield{author}{%
  \bibinfo {author} {\bibfnamefont{A.~W.}\ \bibnamefont{Sandvik}},\ }%
  \bibfield{journal}{%
  \Doi{10.1103/PhysRevE.68.056701}{\bibinfo {journal} {Phys. Rev. E}}\ }%
  \textbf{\bibinfo {volume} {68}},\ \bibinfo {pages} {056701} (\bibinfo {year}
  {2003})%
  \bibAnnoteFile{NoStop}{sandvik_stochastic_2003}%
\bibitem{porras_effective_2004}%
  \BibitemOpen
  \bibfield{author}{%
  \bibinfo {author} {\bibfnamefont{D.}~\bibnamefont{Porras}}\ and\ \bibinfo
  {author} {\bibfnamefont{J.~I.}\ \bibnamefont{Cirac}},\ }%
  \bibfield{journal}{%
  \Doi{10.1103/PhysRevLett.92.207901}{\bibinfo {journal} {Phys. Rev. Lett.}}\
  }%
  \textbf{\bibinfo {volume} {92}},\ \bibinfo {pages} {207901} (\bibinfo {year}
  {2004})%
  \bibAnnoteFile{NoStop}{porras_effective_2004}%
\bibitem{deng_effective_2005}%
  \BibitemOpen
  \bibfield{author}{%
  \bibinfo {author} {\bibfnamefont{X.-L.}\ \bibnamefont{Deng}}, \bibinfo
  {author} {\bibfnamefont{D.}~\bibnamefont{Porras}},\ and\ \bibinfo {author}
  {\bibfnamefont{J.~I.}\ \bibnamefont{Cirac}},\ }%
  \bibfield{journal}{%
  \Doi{10.1103/PhysRevA.72.063407}{\bibinfo {journal} {Phys. Rev. A}}\ }%
  \textbf{\bibinfo {volume} {72}},\ \bibinfo {pages} {063407} (\bibinfo {year}
  {2005})%
  \bibAnnoteFile{NoStop}{deng_effective_2005}%
\bibitem{friedenauer_simulating_2008}%
  \BibitemOpen
  \bibfield{author}{%
  \bibinfo {author} {\bibfnamefont{A.}~\bibnamefont{Friedenauer}}, \bibinfo
  {author} {\bibfnamefont{H.}~\bibnamefont{Schmitz}}, \bibinfo {author}
  {\bibfnamefont{J.~T.}\ \bibnamefont{Glueckert}}, \bibinfo {author}
  {\bibfnamefont{D.}~\bibnamefont{Porras}},\ and\ \bibinfo {author}
  {\bibfnamefont{T.}~\bibnamefont{Schaetz}},\ }%
  \bibfield{journal}{%
  \Doi{10.1038/nphys1032}{\bibinfo {journal} {Nat. Phys.}}\ }%
  \textbf{\bibinfo {volume} {4}},\ \bibinfo {pages} {757} (\bibinfo {year}
  {2008})%
  \bibAnnoteFile{NoStop}{friedenauer_simulating_2008}%
\bibitem{islam_onset_2011}%
  \BibitemOpen
  \bibfield{author}{%
  \bibinfo {author} {\bibfnamefont{R.}~\bibnamefont{Islam}}, \bibinfo {author}
  {\bibfnamefont{E.}~\bibnamefont{Edwards}}, \bibinfo {author}
  {\bibfnamefont{K.}~\bibnamefont{Kim}}, \bibinfo {author}
  {\bibfnamefont{S.}~\bibnamefont{Korenblit}}, \bibinfo {author}
  {\bibfnamefont{C.}~\bibnamefont{Noh}}, \bibinfo {author}
  {\bibfnamefont{H.}~\bibnamefont{Carmichael}}, \bibinfo {author}
  {\bibfnamefont{G.-D.}\ \bibnamefont{Lin}}, \bibinfo {author}
  {\bibfnamefont{L.-M.}\ \bibnamefont{Duan}}, \bibinfo {author}
  {\bibfnamefont{C.-C.}\ \bibnamefont{Joseph~Wang}}, \bibinfo {author}
  {\bibfnamefont{J.}~\bibnamefont{Freericks}},\ and\ \bibinfo {author}
  {\bibfnamefont{C.}~\bibnamefont{Monroe}},\ }%
  \bibfield{journal}{%
  \Doi{10.1038/ncomms1374}{\bibinfo {journal} {Nat. Commun.}}\ }%
  \textbf{\bibinfo {volume} {2}},\ \bibinfo {pages} {377} (\bibinfo {year}
  {2011})%
  \bibAnnoteFile{NoStop}{islam_onset_2011}%
\bibitem{kim_entanglement_2009}%
  \BibitemOpen
  \bibfield{author}{%
  \bibinfo {author} {\bibfnamefont{K.}~\bibnamefont{Kim}}, \bibinfo {author}
  {\bibfnamefont{M.-S.}\ \bibnamefont{Chang}}, \bibinfo {author}
  {\bibfnamefont{R.}~\bibnamefont{Islam}}, \bibinfo {author}
  {\bibfnamefont{S.}~\bibnamefont{Korenblit}}, \bibinfo {author}
  {\bibfnamefont{L.-M.}\ \bibnamefont{Duan}},\ and\ \bibinfo {author}
  {\bibfnamefont{C.}~\bibnamefont{Monroe}},\ }%
  \bibfield{journal}{%
  \Doi{10.1103/PhysRevLett.103.120502}{\bibinfo {journal} {Phys. Rev. Lett.}}\
  }%
  \textbf{\bibinfo {volume} {103}},\ \bibinfo {pages} {120502} (\bibinfo {year}
  {2009})%
  \bibAnnoteFile{NoStop}{kim_entanglement_2009}%
\bibitem{kim_quantum_2010}%
  \BibitemOpen
  \bibfield{author}{%
  \bibinfo {author} {\bibfnamefont{K.}~\bibnamefont{Kim}}, \bibinfo {author}
  {\bibfnamefont{M.-S.}\ \bibnamefont{Chang}}, \bibinfo {author}
  {\bibfnamefont{S.}~\bibnamefont{Korenblit}}, \bibinfo {author}
  {\bibfnamefont{R.}~\bibnamefont{Islam}}, \bibinfo {author}
  {\bibfnamefont{E.~E.}\ \bibnamefont{Edwards}}, \bibinfo {author}
  {\bibfnamefont{J.~K.}\ \bibnamefont{Freericks}}, \bibinfo {author}
  {\bibfnamefont{G.-D.}\ \bibnamefont{Lin}}, \bibinfo {author}
  {\bibfnamefont{L.-M.}\ \bibnamefont{Duan}},\ and\ \bibinfo {author}
  {\bibfnamefont{C.}~\bibnamefont{Monroe}},\ }%
  \bibfield{journal}{%
  \Doi{10.1038/nature09071}{\bibinfo {journal} {Nature}}\ }%
  \textbf{\bibinfo {volume} {465}},\ \bibinfo {pages} {590} (\bibinfo {year}
  {2010})%
  \bibAnnoteFile{NoStop}{kim_quantum_2010}%
\bibitem{edwards_quantum_2010}%
  \BibitemOpen
  \bibfield{author}{%
  \bibinfo {author} {\bibfnamefont{E.~E.}\ \bibnamefont{Edwards}}, \bibinfo
  {author} {\bibfnamefont{S.}~\bibnamefont{Korenblit}}, \bibinfo {author}
  {\bibfnamefont{K.}~\bibnamefont{Kim}}, \bibinfo {author}
  {\bibfnamefont{R.}~\bibnamefont{Islam}}, \bibinfo {author}
  {\bibfnamefont{M.-S.}\ \bibnamefont{Chang}}, \bibinfo {author}
  {\bibfnamefont{J.~K.}\ \bibnamefont{Freericks}}, \bibinfo {author}
  {\bibfnamefont{G.-D.}\ \bibnamefont{Lin}}, \bibinfo {author}
  {\bibfnamefont{L.-M.}\ \bibnamefont{Duan}},\ and\ \bibinfo {author}
  {\bibfnamefont{C.}~\bibnamefont{Monroe}},\ }%
  \bibfield{journal}{%
  \Doi{10.1103/PhysRevB.82.060412}{\bibinfo {journal} {Phys. Rev. B}}\ }%
  \textbf{\bibinfo {volume} {82}},\ \bibinfo {pages} {060412(R)} (\bibinfo
  {year} {2010})%
  \bibAnnoteFile{NoStop}{edwards_quantum_2010}%
\bibitem{lanyon_universal_2011}%
  \BibitemOpen
  \bibfield{author}{%
  \bibinfo {author} {\bibfnamefont{B.~P.}\ \bibnamefont{Lanyon}}, \bibinfo
  {author} {\bibfnamefont{C.}~\bibnamefont{Hempel}}, \bibinfo {author}
  {\bibfnamefont{D.}~\bibnamefont{Nigg}}, \bibinfo {author}
  {\bibfnamefont{M.}~\bibnamefont{M{\"u}ller}}, \bibinfo {author}
  {\bibfnamefont{R.}~\bibnamefont{Gerritsma}}, \bibinfo {author}
  {\bibfnamefont{F.}~\bibnamefont{Z{\"a}hringer}}, \bibinfo {author}
  {\bibfnamefont{P.}~\bibnamefont{Schindler}}, \bibinfo {author}
  {\bibfnamefont{J.~T.}\ \bibnamefont{Barreiro}}, \bibinfo {author}
  {\bibfnamefont{M.}~\bibnamefont{Rambach}}, \bibinfo {author}
  {\bibfnamefont{G.}~\bibnamefont{Kirchmair}}, \bibinfo {author}
  {\bibfnamefont{M.}~\bibnamefont{Hennrich}}, \bibinfo {author}
  {\bibfnamefont{P.}~\bibnamefont{Zoller}}, \bibinfo {author}
  {\bibfnamefont{R.}~\bibnamefont{Blatt}},\ and\ \bibinfo {author}
  {\bibfnamefont{C.~F.}\ \bibnamefont{Roos}},\ }%
  \bibfield{journal}{%
  \Doi{10.1126/science.1208001}{\bibinfo {journal} {Science}}\ }%
  \textbf{\bibinfo {volume} {334}},\ \bibinfo {pages} {57 } (\bibinfo {year}
  {2011})%
  \bibAnnoteFile{NoStop}{lanyon_universal_2011}%
\bibitem{britton_engineered_2012}%
  \BibitemOpen
  \bibfield{author}{%
  \bibinfo {author} {\bibfnamefont{J.~W.}\ \bibnamefont{Britton}}, \bibinfo
  {author} {\bibfnamefont{B.~C.}\ \bibnamefont{Sawyer}}, \bibinfo {author}
  {\bibfnamefont{A.~C.}\ \bibnamefont{Keith}}, \bibinfo {author}
  {\bibfnamefont{C.-C.~J.}\ \bibnamefont{Wang}}, \bibinfo {author}
  {\bibfnamefont{J.~K.}\ \bibnamefont{Freericks}}, \bibinfo {author}
  {\bibfnamefont{H.}~\bibnamefont{Uys}}, \bibinfo {author}
  {\bibfnamefont{M.~J.}\ \bibnamefont{Biercuk}},\ and\ \bibinfo {author}
  {\bibfnamefont{J.~J.}\ \bibnamefont{Bollinger}},\ }%
  \bibfield{journal}{%
  \Doi{10.1038/nature10981}{\bibinfo {journal} {Nature}}\ }%
  \textbf{\bibinfo {volume} {484}},\ \bibinfo {pages} {489} (\bibinfo {year}
  {2012})%
  \bibAnnoteFile{NoStop}{britton_engineered_2012}%
\bibitem{islam_emergence_2013}%
  \BibitemOpen
  \bibfield{author}{%
  \bibinfo {author} {\bibfnamefont{R.}~\bibnamefont{Islam}}, \bibinfo {author}
  {\bibfnamefont{C.}~\bibnamefont{Senko}}, \bibinfo {author}
  {\bibfnamefont{W.~C.}\ \bibnamefont{Campbell}}, \bibinfo {author}
  {\bibfnamefont{S.}~\bibnamefont{Korenblit}}, \bibinfo {author}
  {\bibfnamefont{J.}~\bibnamefont{Smith}}, \bibinfo {author}
  {\bibfnamefont{A.}~\bibnamefont{Lee}}, \bibinfo {author}
  {\bibfnamefont{E.~E.}\ \bibnamefont{Edwards}}, \bibinfo {author}
  {\bibfnamefont{C.-C.~J.}\ \bibnamefont{Wang}}, \bibinfo {author}
  {\bibfnamefont{J.~K.}\ \bibnamefont{Freericks}},\ and\ \bibinfo {author}
  {\bibfnamefont{C.}~\bibnamefont{Monroe}},\ }%
  \bibfield{journal}{%
  \Doi{10.1126/science.1232296}{\bibinfo {journal} {Science}}\ }%
  \textbf{\bibinfo {volume} {340}},\ \bibinfo {pages} {583} (\bibinfo {year}
  {2013})%
  \bibAnnoteFile{NoStop}{islam_emergence_2013}%
\bibitem{richerme_2013}%
  \BibitemOpen
  \bibfield{author}{%
  \bibinfo {author} {\bibfnamefont{P.}~\bibnamefont{Richerme}}, \bibinfo
  {author} {\bibfnamefont{C.}~\bibnamefont{Senko}}, \bibinfo {author}
  {\bibfnamefont{S.}~\bibnamefont{Korenblit}}, \bibinfo {author}
  {\bibfnamefont{J.}~\bibnamefont{Smith}}, \bibinfo {author}
  {\bibfnamefont{A.}~\bibnamefont{Lee}}, \bibinfo {author}
  {\bibfnamefont{R.}~\bibnamefont{Islam}}, \bibinfo {author}
  {\bibfnamefont{W.~C.}\ \bibnamefont{Campbell}},\ and\ \bibinfo {author}
  {\bibfnamefont{C.}~\bibnamefont{Monroe}},\ }%
  \bibfield{journal}{%
  \Doi{10.1103/PhysRevLett.111.100506}{\bibinfo {journal} {Phys. Rev. Lett.}}\
  }%
  \textbf{\bibinfo {volume} {111}},\ \bibinfo {pages} {100506} (\bibinfo {year}
  {2013})%
  \bibAnnoteFile{NoStop}{richerme_2013}%
\bibitem{laflorencie_critical_2005}%
  \BibitemOpen
  \bibfield{author}{%
  \bibinfo {author} {\bibfnamefont{N.}~\bibnamefont{Laflorencie}}, \bibinfo
  {author} {\bibfnamefont{I.}~\bibnamefont{Affleck}},\ and\ \bibinfo {author}
  {\bibfnamefont{M.}~\bibnamefont{Berciu}},\ }%
  \bibfield{journal}{%
  \Doi{10.1088/1742-5468/2005/12/P12001}{\bibinfo {journal} {J. Stat. Mech.}}\
  }%
  \textbf{\bibinfo {volume} {2005}},\ \bibinfo {pages} {P12001} (\bibinfo
  {year} {2005})%
  \bibAnnoteFile{NoStop}{laflorencie_critical_2005}%
\bibitem{sandvik_ground_2010}%
  \BibitemOpen
  \bibfield{author}{%
  \bibinfo {author} {\bibfnamefont{A.~W.}\ \bibnamefont{Sandvik}},\ }%
  \bibfield{journal}{%
  \Doi{10.1103/PhysRevLett.104.137204}{\bibinfo {journal} {Phys. Rev. Lett.}}\
  }%
  \textbf{\bibinfo {volume} {104}},\ \bibinfo {pages} {137204} (\bibinfo {year}
  {2010})%
  \bibAnnoteFile{NoStop}{sandvik_ground_2010}%
\bibitem{koffel_entanglement_2012}%
  \BibitemOpen
  \bibfield{author}{%
  \bibinfo {author} {\bibfnamefont{T.}~\bibnamefont{Koffel}}, \bibinfo {author}
  {\bibfnamefont{M.}~\bibnamefont{Lewenstein}},\ and\ \bibinfo {author}
  {\bibfnamefont{L.}~\bibnamefont{Tagliacozzo}},\ }%
  \bibfield{journal}{%
  \Doi{10.1103/PhysRevLett.109.267203}{\bibinfo {journal} {Phys. Rev. Lett.}}\
  }%
  \textbf{\bibinfo {volume} {109}},\ \bibinfo {pages} {267203} (\bibinfo {year}
  {2012})%
  \bibAnnoteFile{NoStop}{koffel_entanglement_2012}%
\bibitem{wall_out_2012}%
  \BibitemOpen
  \bibfield{author}{%
  \bibinfo {author} {\bibfnamefont{M.~L.}\ \bibnamefont{Wall}}\ and\ \bibinfo
  {author} {\bibfnamefont{L.~D.}\ \bibnamefont{Carr}},\ }%
  \bibfield{journal}{%
  \Doi{10.1088/1367-2630/14/12/125015}{\bibinfo {journal} {New J. Phys.}}\ }%
  \textbf{\bibinfo {volume} {14}},\ \bibinfo {pages} {125015} (\bibinfo {year}
  {2012})%
  \bibAnnoteFile{NoStop}{wall_out_2012}%
\bibitem{hauke_spread_2013}%
  \BibitemOpen
  \bibfield{author}{%
  \bibinfo {author} {\bibfnamefont{P.}~\bibnamefont{Hauke}}\ and\ \bibinfo
  {author} {\bibfnamefont{L.}~\bibnamefont{Tagliacozzo}},\ }%
  \bibinfo {journal} {arXiv:1304.7725}%
  \bibAnnoteFile{NoStop}{hauke_spread_2013}%
\bibitem{schachenmayer_entanglement_2013}%
  \BibitemOpen
\bibfield{journal}{%
    }%
  \bibfield{author}{%
  \bibinfo {author} {\bibfnamefont{J.}~\bibnamefont{Schachenmayer}}, \bibinfo
  {author} {\bibfnamefont{B.~P.}\ \bibnamefont{Lanyon}}, \bibinfo {author}
  {\bibfnamefont{C.~F.}\ \bibnamefont{Roos}},\ and\ \bibinfo {author}
  {\bibfnamefont{A.~J.}\ \bibnamefont{Daley}},\ }%
  \bibfield{journal}{%
  \Doi{10.1103/PhysRevX.3.031015}{\bibinfo {journal} {Phys. Rev. X}}\ }%
  \textbf{\bibinfo {volume} {3}},\ \bibinfo {pages} {031015} (\bibinfo {year}
  {2013})%
  \bibAnnoteFile{NoStop}{schachenmayer_entanglement_2013}%
\bibitem{saad_numerical_1995}%
  \BibitemOpen
  \bibfield{author}{%
  \bibinfo {author} {\bibfnamefont{Y.}~\bibnamefont{Saad}},\ }%
  \emph{\bibinfo {title} {Numerical Methods for Large Eigenvalue Problems}}\
  (\bibinfo {publisher} {Manchester University Press},\ \bibinfo {year}
  {1995})%
  \bibAnnoteFile{NoStop}{saad_numerical_1995}%
\bibitem{binder_critical_1981}%
  \BibitemOpen
  \bibfield{author}{%
  \bibinfo {author} {\bibfnamefont{K.}~\bibnamefont{Binder}},\ }%
  \bibfield{journal}{%
  \Doi{10.1103/PhysRevLett.47.693}{\bibinfo {journal} {Phys. Rev. Lett.}}\ }%
  \textbf{\bibinfo {volume} {47}},\ \bibinfo {pages} {693} (\bibinfo {year}
  {1981})%
  \bibAnnoteFile{NoStop}{binder_critical_1981}%
\bibitem{basko_metalinsulator_2006}%
  \BibitemOpen
  \bibfield{author}{%
  \bibinfo {author} {\bibfnamefont{D.}~\bibnamefont{Basko}}, \bibinfo {author}
  {\bibfnamefont{I.}~\bibnamefont{Aleiner}},\ and\ \bibinfo {author}
  {\bibfnamefont{B.}~\bibnamefont{Altshuler}},\ }%
  \bibfield{journal}{%
  \Doi{10.1016/j.aop.2005.11.014}{\bibinfo {journal} {Ann. Phys.}}\ }%
  \textbf{\bibinfo {volume} {321}},\ \bibinfo {pages} {1126} (\bibinfo {year}
  {2006})%
  \bibAnnoteFile{NoStop}{basko_metalinsulator_2006}%
\bibitem{gornyi_interacting_2005}%
  \BibitemOpen
  \bibfield{author}{%
  \bibinfo {author} {\bibfnamefont{I.~V.}\ \bibnamefont{Gornyi}}, \bibinfo
  {author} {\bibfnamefont{A.~D.}\ \bibnamefont{Mirlin}},\ and\ \bibinfo
  {author} {\bibfnamefont{D.~G.}\ \bibnamefont{Polyakov}},\ }%
  \bibfield{journal}{%
  \Doi{10.1103/PhysRevLett.95.206603}{\bibinfo {journal} {Phys. Rev. Lett.}}\
  }%
  \textbf{\bibinfo {volume} {95}},\ \bibinfo {pages} {206603} (\bibinfo {year}
  {2005})%
  \bibAnnoteFile{NoStop}{gornyi_interacting_2005}%
\bibitem{oganesyan_localization_2007}%
  \BibitemOpen
  \bibfield{author}{%
  \bibinfo {author} {\bibfnamefont{V.}~\bibnamefont{Oganesyan}}\ and\ \bibinfo
  {author} {\bibfnamefont{D.~A.}\ \bibnamefont{Huse}},\ }%
  \bibfield{journal}{%
  \Doi{10.1103/PhysRevB.75.155111}{\bibinfo {journal} {Phys. Rev. B}}\ }%
  \textbf{\bibinfo {volume} {75}},\ \bibinfo {pages} {155111} (\bibinfo {year}
  {2007})%
  \bibAnnoteFile{NoStop}{oganesyan_localization_2007}%
\bibitem{serbyn_universal_2013}%
  \BibitemOpen
  \bibfield{author}{%
  \bibinfo {author} {\bibfnamefont{M.}~\bibnamefont{Serbyn}}, \bibinfo {author}
  {\bibfnamefont{Z.}~\bibnamefont{Papi{\'c}}},\ and\ \bibinfo {author}
  {\bibfnamefont{D.~A.}\ \bibnamefont{Abanin}},\ }%
  \bibfield{journal}{%
  \Doi{10.1103/PhysRevLett.110.260601}{\bibinfo {journal} {Phys. Rev. Lett.}}\
  }%
  \textbf{\bibinfo {volume} {110}},\ \bibinfo {pages} {260601} (\bibinfo {year}
  {2013})%
  \bibAnnoteFile{NoStop}{serbyn_universal_2013}%
\bibitem{huse_phenomenology_2013}%
  \BibitemOpen
  \bibfield{author}{%
  \bibinfo {author} {\bibfnamefont{D.~A.}\ \bibnamefont{Huse}}\ and\ \bibinfo
  {author} {\bibfnamefont{V.}~\bibnamefont{Oganesyan}},\ }%
  \bibinfo {journal} {{arXiv:1305.4915}}%
  \bibAnnoteFile{NoStop}{huse_phenomenology_2013}%
\bibitem{serbyn_local_2013}%
  \BibitemOpen
\bibfield{journal}{%
    }%
  \bibfield{author}{%
  \bibinfo {author} {\bibfnamefont{M.}~\bibnamefont{Serbyn}}, \bibinfo {author}
  {\bibfnamefont{Z.}~\bibnamefont{Papi{\'c}}},\ and\ \bibinfo {author}
  {\bibfnamefont{D.~A.}\ \bibnamefont{Abanin}},\ }%
  \bibfield{journal}{%
  \Doi{10.1103/PhysRevLett.111.127201}{\bibinfo {journal} {Phys. Rev. Lett.}}\
  }%
  \textbf{\bibinfo {volume} {111}},\ \bibinfo {pages} {127201} (\bibinfo {year}
  {2013})%
  \bibAnnoteFile{NoStop}{serbyn_local_2013}%
\bibitem{rigol_relaxation_2007}%
  \BibitemOpen
  \bibfield{author}{%
  \bibinfo {author} {\bibfnamefont{M.}~\bibnamefont{Rigol}}, \bibinfo {author}
  {\bibfnamefont{V.}~\bibnamefont{Dunjko}}, \bibinfo {author}
  {\bibfnamefont{V.}~\bibnamefont{Yurovsky}},\ and\ \bibinfo {author}
  {\bibfnamefont{M.}~\bibnamefont{Olshanii}},\ }%
  \bibfield{journal}{%
  \Doi{10.1103/PhysRevLett.98.050405}{\bibinfo {journal} {Phys. Rev. Lett.}}\
  }%
  \textbf{\bibinfo {volume} {98}},\ \bibinfo {pages} {050405} (\bibinfo {year}
  {2007})%
  \bibAnnoteFile{NoStop}{rigol_relaxation_2007}%
\bibitem{rigol_thermalization_2008}%
  \BibitemOpen
  \bibfield{author}{%
  \bibinfo {author} {\bibfnamefont{M.}~\bibnamefont{Rigol}}, \bibinfo {author}
  {\bibfnamefont{V.}~\bibnamefont{Dunjko}},\ and\ \bibinfo {author}
  {\bibfnamefont{M.}~\bibnamefont{Olshanii}},\ }%
  \bibfield{journal}{%
  \Doi{10.1038/nature06838}{\bibinfo {journal} {Nature}}\ }%
  \textbf{\bibinfo {volume} {452}},\ \bibinfo {pages} {854} (\bibinfo {year}
  {2008})%
  \bibAnnoteFile{NoStop}{rigol_thermalization_2008}%
\bibitem{simon_quantum_2011}%
  \BibitemOpen
  \bibfield{author}{%
  \bibinfo {author} {\bibfnamefont{J.}~\bibnamefont{Simon}}, \bibinfo {author}
  {\bibfnamefont{W.~S.}\ \bibnamefont{Bakr}}, \bibinfo {author}
  {\bibfnamefont{R.}~\bibnamefont{Ma}}, \bibinfo {author}
  {\bibfnamefont{M.~E.}\ \bibnamefont{Tai}}, \bibinfo {author}
  {\bibfnamefont{P.~M.}\ \bibnamefont{Preiss}},\ and\ \bibinfo {author}
  {\bibfnamefont{M.}~\bibnamefont{Greiner}},\ }%
  \bibfield{journal}{%
  \Doi{10.1038/nature09994}{\bibinfo {journal} {Nature}}\ }%
  \textbf{\bibinfo {volume} {472}},\ \bibinfo {pages} {307} (\bibinfo {year}
  {2011})%
  \bibAnnoteFile{NoStop}{simon_quantum_2011}%
\bibitem{struck_quantum_2011}%
  \BibitemOpen
  \bibfield{author}{%
  \bibinfo {author} {\bibfnamefont{J.}~\bibnamefont{Struck}}, \bibinfo {author}
  {\bibfnamefont{C.}~\bibnamefont{{\"O}lschl{\"a}ger}}, \bibinfo {author}
  {\bibfnamefont{R.~L.}\ \bibnamefont{Targat}}, \bibinfo {author}
  {\bibfnamefont{P.}~\bibnamefont{Soltan-Panahi}}, \bibinfo {author}
  {\bibfnamefont{A.}~\bibnamefont{Eckardt}}, \bibinfo {author}
  {\bibfnamefont{M.}~\bibnamefont{Lewenstein}}, \bibinfo {author}
  {\bibfnamefont{P.}~\bibnamefont{Windpassinger}},\ and\ \bibinfo {author}
  {\bibfnamefont{K.}~\bibnamefont{Sengstock}},\ }%
  \bibfield{journal}{%
  \Doi{10.1126/science.1207239}{\bibinfo {journal} {Science}}\ }%
  \textbf{\bibinfo {volume} {333}},\ \bibinfo {pages} {996} (\bibinfo {year}
  {2011})%
  \bibAnnoteFile{NoStop}{struck_quantum_2011}%
\bibitem{thouless_long-range_1969}%
  \BibitemOpen
  \bibfield{author}{%
  \bibinfo {author} {\bibfnamefont{D.~J.}\ \bibnamefont{Thouless}},\ }%
  \bibfield{journal}{%
  \Doi{10.1103/PhysRev.187.732}{\bibinfo {journal} {Phys. Rev.}}\ }%
  \textbf{\bibinfo {volume} {187}},\ \bibinfo {pages} {732} (\bibinfo {year}
  {1969})%
  \bibAnnoteFile{NoStop}{thouless_long-range_1969}%
\bibitem{kosterlitz_phase_1976}%
  \BibitemOpen
  \bibfield{author}{%
  \bibinfo {author} {\bibfnamefont{J.~M.}\ \bibnamefont{Kosterlitz}},\ }%
  \bibfield{journal}{%
  \Doi{10.1103/PhysRevLett.37.1577}{\bibinfo {journal} {Phys. Rev. Lett.}}\ }%
  \textbf{\bibinfo {volume} {37}},\ \bibinfo {pages} {1577} (\bibinfo {year}
  {1976})%
  \bibAnnoteFile{NoStop}{kosterlitz_phase_1976}%
\bibitem{cardy_one-dimensional_1981}%
  \BibitemOpen
  \bibfield{author}{%
  \bibinfo {author} {\bibfnamefont{J.~L.}\ \bibnamefont{Cardy}},\ }%
  \bibfield{journal}{%
  \Doi{10.1088/0305-4470/14/6/017}{\bibinfo {journal} {J. Phys. A: Math.
  Gen.}}\ }%
  \textbf{\bibinfo {volume} {14}},\ \bibinfo {pages} {1407} (\bibinfo {year}
  {1981})%
  \bibAnnoteFile{NoStop}{cardy_one-dimensional_1981}%
\bibitem{bhattacharjee_properties_1981}%
  \BibitemOpen
  \bibfield{author}{%
  \bibinfo {author} {\bibfnamefont{J.}~\bibnamefont{Bhattacharjee}}, \bibinfo
  {author} {\bibfnamefont{S.}~\bibnamefont{Chakravarty}}, \bibinfo {author}
  {\bibfnamefont{J.~L.}\ \bibnamefont{Richardson}},\ and\ \bibinfo {author}
  {\bibfnamefont{D.~J.}\ \bibnamefont{Scalapino}},\ }%
  \bibfield{journal}{%
  \Doi{10.1103/PhysRevB.24.3862}{\bibinfo {journal} {Phys. Rev. B}}\ }%
  \textbf{\bibinfo {volume} {24}},\ \bibinfo {pages} {3862} (\bibinfo {year}
  {1981})%
  \bibAnnoteFile{NoStop}{bhattacharjee_properties_1981}%
\bibitem{jaklic_lanczos_1994}%
  \BibitemOpen
  \bibfield{author}{%
  \bibinfo {author} {\bibfnamefont{J.}~\bibnamefont{Jakli{\v c}}}\ and\
  \bibinfo {author} {\bibfnamefont{P.}~\bibnamefont{Prelov{\v s}ek}},\ }%
  \bibfield{journal}{%
  \Doi{10.1103/PhysRevB.49.5065}{\bibinfo {journal} {Phys. Rev. B}}\ }%
  \textbf{\bibinfo {volume} {49}},\ \bibinfo {pages} {5065} (\bibinfo {year}
  {1994})%
  \bibAnnoteFile{NoStop}{jaklic_lanczos_1994}%
\bibitem{aichhorn_low-temperature_2003}%
  \BibitemOpen
  \bibfield{author}{%
  \bibinfo {author} {\bibfnamefont{M.}~\bibnamefont{Aichhorn}}, \bibinfo
  {author} {\bibfnamefont{M.}~\bibnamefont{Daghofer}}, \bibinfo {author}
  {\bibfnamefont{H.~G.}\ \bibnamefont{Evertz}},\ and\ \bibinfo {author}
  {\bibfnamefont{W.}~\bibnamefont{von~der Linden}},\ }%
  \bibfield{journal}{%
  \Doi{10.1103/PhysRevB.67.161103}{\bibinfo {journal} {Phys. Rev. B}}\ }%
  \textbf{\bibinfo {volume} {67}},\ \bibinfo {pages} {161103(R)} (\bibinfo
  {year} {2003})%
  \bibAnnoteFile{NoStop}{aichhorn_low-temperature_2003}%
\end{thebibliography}
%


\appendix

\section{Technical details on many-body Ramsey interferometry}

We present the calculation of the full expectation value \eqw{eq:Mexp}, including 
the odd terms of spin operators, which is measured with many-body Ramsey interferometry.
To this end we introduce
\begin{align*}
 \hat C(t) :=& e^{i {\hat H} t} R^\dagger(\phi_2)
\sigma_{j}^{z} R(\phi_2) e^{- i {\hat H} t} \\=& - [\sigma_j^x(t) \sin \phi_2 + \sigma_j^y(t) \cos \phi_2]\;.
\end{align*}
With that we obtain 
\begin{align*}
  M_{ij}& = \frac{1}{2} \langle i [\hat C(t),\sigma_i^x] \cos \phi_1 - i [\hat C(t),\sigma_i^y] \sin \phi_1+ \hat C(t) \\
& + (\sigma_i^x \cos \phi_1 - \sigma_i^y \sin \phi_1)\hat C(t)(\sigma_i^x \cos \phi_1 - \sigma_i^y \sin \phi_1)\rangle\;.
\end{align*}
The commutators of $\hat C(t)$ with the spin operators $\sigma_i^{x,y}$ yield
the Green's functions, whereas the other terms contain either one or three spin operators.

As discussed in the main text, the expectation value of odd numbers of spin operators 
vanishes trivially for the Heisenberg model. For the long-ranged, transverse field
Ising model they vanish provided
\[
 \cos \phi_2 =0 \quad \text{and} \quad \sin \phi_1 \cos \phi_1 = 0 \;,
\]
which is fulfilled when locking the phases of the laser field to $\phi_2=\pi/2$ and 
$\phi_1=\lbrace  0, \pi/2 \rbrace$. For the latter choice of the phase $\phi_1=\pi/2$ 
the whole expression \eqw{eq:Mexp} vanishes, whereas for $\phi_1=0$ it gives \eq{eq:Mising}.

\section{Spin-shelving protocol}

\begin{figure}
\begin{center}
 \includegraphics[width=0.3\textwidth]{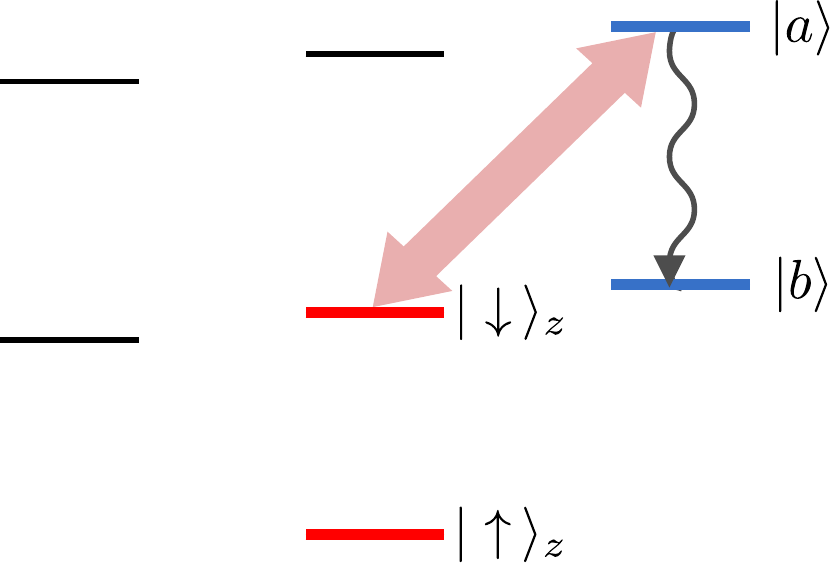}
\end{center}
\caption{\label{fig:spLevels} (Color online) Level scheme required for the spin-shelving protocol. 
The thick red arrow illustrates the strong drive between $\d_z$ and an auxiliary level $\ket{a}$, 
which decays spontaneously to a metastable state $\ket{b}$, as indicated by the black wavy arrow
(see text for details).
}
\end{figure}

Projecting one spin state onto an auxiliary level, allows one to
extract the anticommutator,  Green's function $G^{zz,+}_{ij}$ in the orthogonal spin direction compared
to the Green's functions accessible via many-body Ramsey interference. The protocol is as follows: 
First, a transition between one state of the two level system, say $\d_z$, 
and an auxiliary level $\ket{a}$  has to be driven strongly, see \fig{fig:spLevels} for the level scheme. The state $\ket{a}$ 
should decay spontaneously into a metastable state $\ket{b}$. This 
process shelves the $\d_z$ state to the auxiliary 
level $\ket{b}$, which might give rise to a complicated time 
evolution. However, the protocol includes disregarding all experimental 
runs where auxiliary level $\ket{b}$ is populated at the end of the time evolution. Following that 
protocol a spin projection operator of the form $\hat P_i = (\hat{\mathds{1}}+\sigma^z_i)/2$ 
is realized which allows one to extract the anticommutator Green's function
\begin{align}
 &\langle \hat P_i^\dag e^{i \hat H t} \sigma_j^z  e^{-i \hat H t} \hat P_i \rangle = i G^{zz,+}_{ij} + \text{terms with one } \sigma^{z} \;.
\end{align}
The experimental challenge of the protocol is that instead of 
two states, three states have to be detected. 

For the Heisenberg model \eqw{eq:heis}, the spin-shelving technique measures 
$G^{zz,+}_{ij}$ in the case of zero magnetization $m_z = \sum_r \sigma_r^z =0 $, where 
the ground state has the additional global symmetry $\sigma^x \rightarrow - \sigma^x$, $\sigma^y \rightarrow \sigma^y$ 
and $\sigma^z \rightarrow - \sigma^z$ and thus expectation values with odd numbers
of spin-z operators vanish.

In the case of the long-range, transverse field Ising model~\eqw{eq:ising} that global 
symmetry is trivially fulfilled. Thus the spin-shelving 
technique allows one to measure $G^{zz,+}_{ij}$, which is further related to 
$G^{xx,+}_{ij}(t)$ through 
\begin{equation}
G^{zz,\mp}_{ij}(t)=-\frac{1}{4h^2} \frac{d^2}{dt^2} G^{xx,\mp}_{ij}(t) \;,      
\label{eq:gzgx}
\end{equation}
as commuting $\hat H_{\text{Ising}}$ with $\sigma_i^x$ generates $\sigma_i^z$. Hence, for
the long-ranged, transverse field Ising model \eqw{eq:ising} the spin-shelving 
technique also makes it possible to explore $G^{xx,+}_{ij}(t)$. 

\section{Loschmidt echo}

The Loschmidt echo is defined as
\[
 \mathcal{L}(t) := \frac{1}{Z} \sum_n e^{-\beta E_n} \langle n |  e^{i \hat H_0 t} e^{-i \hat H_1 t} | n  \rangle\;,
\]
which describes a forward propagation in time with Hamiltonian $\hat H_1 := \hat H_0 + \hat V$ and a
backward propagation with $\hat H_0$ for the same time period. In the following, 
we assume that $\hat V$ is a local perturbation.
We expect that if the system is in a localized regime, a local perturbation does not have a dramatic
effect and eigenstates of $\hat H_0+\hat V$ and $\hat H_0$ differ only slightly. 
Hence $\mathcal{L}(t)$ will oscillate in time without fully decaying. 
By contrast, if the system is in a diffusive regime, $\mathcal{L}(t)$ should
decay quickly in time. 

It is useful to point out that Green's functions which are local in space are directly connected 
to the Loschmidt echo
\begin{align*}
 G^{aa,\mp}_{jj} &= -{i \theta(t)}\langle \sigma^a_j(t) \sigma^a_j(0) \mp \sigma^a_j(0) \sigma^a_j(t)  \rangle \nonumber \\
 &= -i \theta(t)\langle  e^{i \hat H t} \underbrace{\sigma^a_j e^{-i \hat H t} \sigma^a_j}_{:=e^{i \hat H_1 t}} \mp {\sigma^a_j e^{i \hat H t}\sigma^a_j} e^{-i \hat H t} \rangle \nonumber \\   &= -i \theta(t) [\mathcal{L}(t) \mp \mathcal{L}(-t)]
\end{align*}
with $\hat H_0 = \hat H$ and $\hat H_1$ is identical to $\hat H$ except for the local spin transformation: $\sigma^a_j \to \sigma^a_j$ 
and $\sigma^b_j \to -\sigma^b_j$, for $b \neq a$.

From these considerations immediately follows that the proposed many-body Ramsey 
interference and the spin-shelving technique can be used
to distinguish localized and diffusive phases.

\section{Spin-echo protocol}

An external magnetic field, which slowly changes between individual experimental runs, couples to 
the system in form of a Zeeman term 
\begin{equation}
{\hat H}_{\rm Z} = h_z \sum_i \sigma^z_i\;.
\label{eq:zee}
\end{equation}
Here we discuss under which requirements spin echo can be used to remove these fluctuations.
The spin-echo protocol differs from the many-body Ramsey protocol by a global $\pi$ rotation $R_\pi:=R(\pi,\phi_\pi)$
performed at time $t/2$
\begin{widetext}
\begin{align}
\tilde M_{ij}( \phi_1,  \phi_2, t) = \sum_n \frac{e^{-\beta E_n}}{Z}
\langle n | R^\dagger_{i}(\phi_1) e^{ i {\hat H} t/2} R_\pi^\dag e^{i {\hat H} t/2} R^\dagger(\phi_2)
\sigma_{j}^{z} R(\phi_2) e^{- i {\hat H} t/2} R_\pi e^{- i {\hat H} t/2} R_{i}(\phi_1) | n \rangle \;.
\end{align}
The rotation $R_\pi$, see \eq{eq:rot}, explicitly reads
\[
 R_\pi = \prod_j i(\sigma_j^x\cos \phi_\pi-\sigma_j^y\sin \phi_\pi)\;.
\]
It transforms the spins of the Hamiltonian as follows:
\begin{equation}
 \forall \text{ sites: }\sigma^a \to \sigma^a, \; \sigma^b \to -\sigma^b, \;\text{for}\;a\neq b \;.
 \label{eq:se}
\end{equation}
Using the fact that $a\in \lbrace x,y\rbrace$ the sign of $\sigma^z$ is always 
flipped under that transformation. Thus 
the Zeeman field \eq{eq:zee} in $z$-direction is canceled, as from 
$0$ to $t/2$ ${\hat H}_{\rm Z}$ evolves with positive sign, whereas the 
from $t/2$ to $t$ its evolution enters with negative sign, provided the Hamiltonian
commutes with $\hat H_\text{Z}$, see also \eq{eq:seShort}.

The Heisenberg model \eqw{eq:heis} fulfills these requirements and thus
Zeeman field fluctuations can be removed using 
the spin-echo protocol. A detailed calculation shows that
\begin{align}
 \tilde M_{ij}( \phi_1,  \phi_2, t) = \frac{1}{4}  \cos 2 \phi_\pi \big \lbrace &\sin (\phi_1 + \phi_2) ( G_{ij}^{xx}  +  G_{ij}^{yy} ) -\cos (\phi_1 + \phi_2) ( G_{ij}^{xy}  -  G_{ij}^{yx} ) \big \rbrace  \nonumber	 \\
  -\frac{1}{4}  \sin 2 \phi_\pi \big \lbrace &\cos (\phi_1 + \phi_2) ( G_{ij}^{xx}  +  G_{ij}^{yy} ) -  \sin (\phi_1 + \phi_2) ( G_{ij}^{xy}  -  G_{ij}^{yx} ) \big \rbrace  \;.
\label{eq:seLong}
 \end{align}
For the choice of $\phi_\pi=0$ or  $\phi_\pi=\pi/2$ only the first line of \eq{eq:seLong} contributes and 
corresponds to the many-body Ramsey interference \eqw{MshortHeisenberg} with $\phi_2\to-\phi_2$.
\end{widetext}

The long-range, transverse field Ising model \eqw{eq:ising}, is only invariant under
\eqw{eq:se} if $a=y$, \ie, $\phi_\pi=\pi/2$. As for many-body Ramsey interference,
the phases $\phi_1$ and $\phi_2$ have to be chosen such that the odd number of $\sigma^{x,y}$
operators vanish. This can be achieved again with $\phi_1=0$ and $\phi_2=\pi/2$ where
we find
\begin{equation}
 \tilde M_{ij}( 0,  \pi/2, t) = -\frac{1}{2}  G^{xx}_{ij} \;.
\end{equation}
However, the Zeeman term does not commute with the Ising Hamiltonian. Thus
the evolution of both contributions is entangled and cannot be undone with
spin-echo. Using the Baker-Campbell-Hausdorff formula, it can be shown that the 
dynamic is governed by the Ising Hamiltonian  to order O$(t^2 h_z)$, \ie, for short times and small changes 
in the magnetic field. However, for current experimental
realizations with trapped ${}^{171}\text{Yb}^+$ ions it is not necessary to 
aim at a spin-echo procedure, as the hyperfine-states which are used to encode 
the spin states do not couple to the Zeeman field.
  
\section{Dynamic signature of long-range order in the Heisenberg antiferromagnet}

In the two-dimensional Heisenberg antiferromagnet long-range order occurs only at
zero temperature. However, even at non-zero temperatures the dynamic structure factor
contains signatures of the long-range order when evaluated at the antiferromagnetic 
ordering wave-vector $\ve{q}=\vec \pi$. The dynamic structure factor is defined as
\begin{equation}
 S^\alpha_{\ve q}(t) := \sum_\ve{r} e^{-i \ve{q} \ve{r}} \langle \sigma^\alpha_{\ve{r}}(t) \sigma^\alpha_{\ve{0}} \rangle \,.
 \label{eq:Sqf}
\end{equation}
and can thus be evaluated from the spatially resolved Ramsey experiments when
adding up contributions from one sublattice with positive sign and the other 
with negative sign.

\begin{figure}
\begin{center}
 \includegraphics[width=0.49\textwidth]{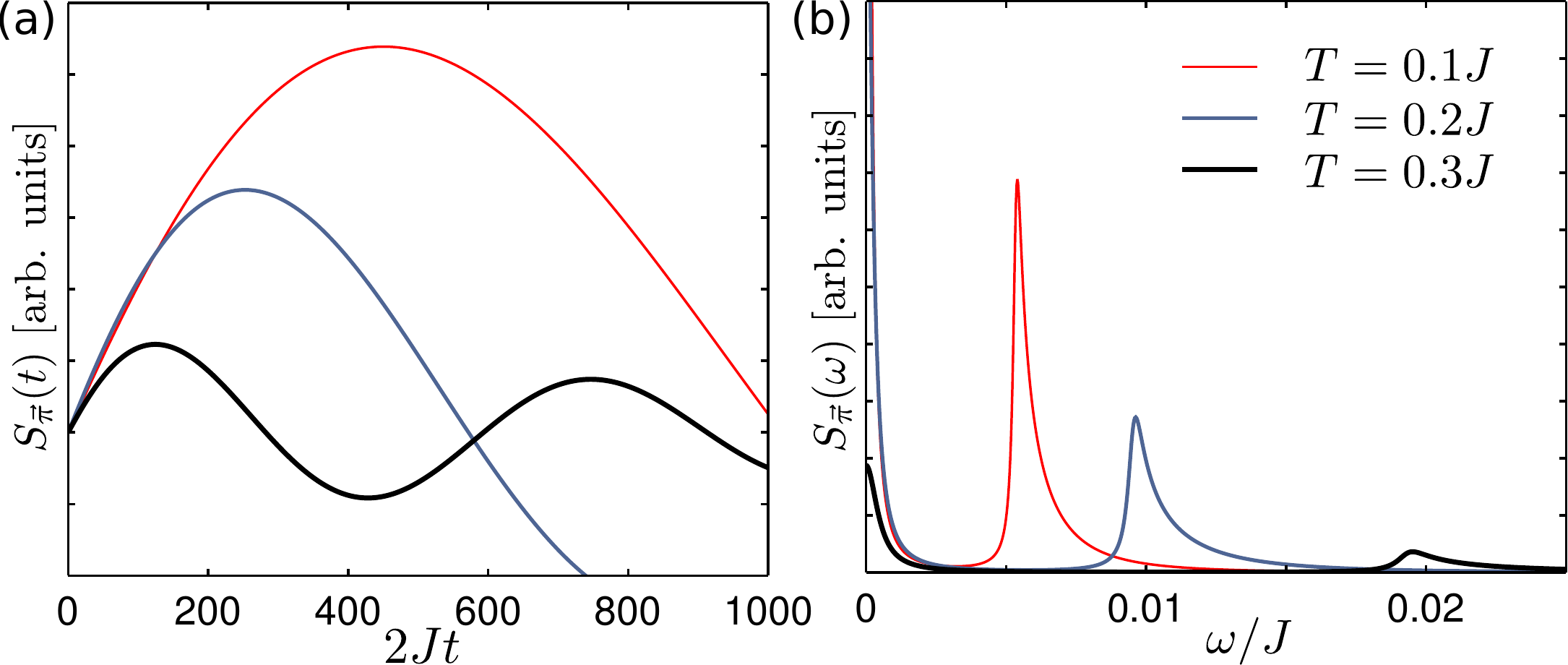}
\end{center}
\caption{\label{fig:hSQ} (Color online) Time dependent \fc{a} and frequency resolved \fc{b} 
structure factor at the antiferromagnetic ordering wave-vector $\vec \pi$ evaluated at different temperatures,
see legend. The built up of antiferromagnetic correlations manifests in the strong increase of the
period and the amplitude of the oscillations in the time dependent structure factor.
}
\end{figure}

Typically, in condensed matter experiments, the dynamic structure factor is measured
as a function of frequency $S^\alpha_{\ve q}(\omega)$, which is related to $S^\alpha_{\ve q}(t)$
through Fourier transformation. A signature of long-range oder in $S^\alpha_{\vec \pi}(\omega)$ 
is mode softening and accumulation of the zero energy peak with decreasing temperature. This is 
demonstrated in \figc{fig:hSQ}{b}. In the time resolved measurement the absence of antiferromagnetic order would
manifest in fast, small-amplitude oscillations of $S_{\vec \pi}(t)$, whereas the onset of
long-range order manifests in a dramatic increase of the amplitude and period of the 
oscillations, as with decreasing temperature smaller energy scales are involved, \figc{fig:hSQ}{a}.

\section{Quantum phase transition of the ferromagnetic, long-range, transverse field Ising model}

In \figc{fig:pd}{a} we show the phase of the one-dimensional, ferromagnetic, long-range, transverse 
field Ising model diagram in the transverse field $h$, 
interaction exponent $\alpha$, and temperature $T$ space, which exhibits particularly rich physics. For $\alpha<1$ the
system is thermodynamically unstable as the energy per site diverges, hatched region. 
Starting out with a Ginzburg-Landau action, we summarize the main properties of the zero temperature phase
transition (see also~\cite{dutta_phase_2001})
\begin{equation}
 S = \frac12 \int \frac{d\omega}{2\pi} \int \frac{dk}{2\pi} (\omega^2 + ck^2+ r+V(k))\phi_k^2  + \frac{u}{4!} S_u
\label{eq:gl}
\end{equation}
where $S_u$ contains the $\phi^4$ term and $V(k)$ is the Fourier transform of 
the long-range interactions which scales as $\sim k^{\alpha-1}$. Thus
$V(k)$ renormalizes to zero for $\alpha\geq 3$ and the phase transition is of the universality 
class of the short-range Ising transition, dark gray region. For $\alpha<3$ the free propagator is of the form
$G^{xx}_{(0)}(k,\omega) \sim (\omega^2+r+v k^{\alpha-1})^{-1}$. From the temporal
and the spatial correlations along the critical line, the dynamical critical 
exponent $z(\alpha)$, which relates the scaling of space and time 
\begin{equation}
 t \sim r^{z(\alpha)} \,
 \label{eq:z}
\end{equation}
can be extracted. $z(\alpha)$ is a continuous function of the interaction exponent $\alpha$.
From the free propagator we find at the critical point $r=0$ 
the mean-field dynamical critical exponent $z^{\text{mf}}=(\alpha-1)/2$, 
and spatial critical exponent $\eta_x^{\text{mf}}=3-\alpha$. Instead of an upper 
critical dimension, we can now talk about a lower critical interaction exponent 
$\alpha_l$ below which mean-field becomes exact. Power counting and applying the condition 
that the scaling dimension of the coefficient $u$ of the $\phi^4$ term [$S_u$ in \eq{eq:gl}] vanishes, 
gives $\alpha_l=5/3$. For interactions which decay slower, as indicated by the light gray region, 
mean-field exponents are valid. 
\begin{figure}
\begin{center}
 \includegraphics[width=0.48\textwidth]{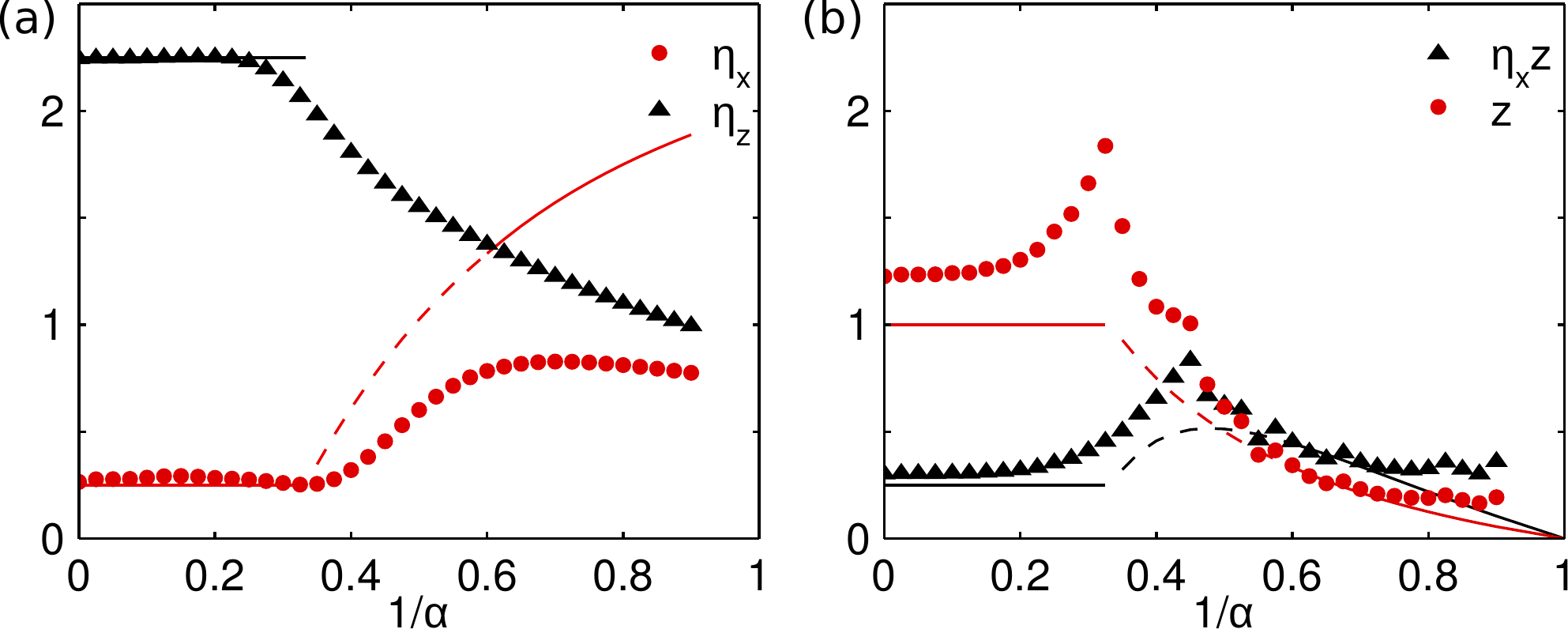}
\end{center}
\caption{\label{fig:qc} (Color online) \fc{a} Critical exponent $\eta_x$ ($\eta_z$), 
symbols, extracted from the extrapolation of finite size data
 of the spatial correlations along the quantum critical line and \fc{b} critical exponent $\eta_x z$ of dynamic correlations $G^{xx,-}_{L/2,L/2}(t)$
and dynamical exponent $z$, symbols. As a function of the interaction exponent $\alpha$ the quantum
phase transition follows a continuous manifold of universality classes characterized
by a set of continuously changing critical exponents. Solid lines are exact results for the critical exponents 
in the thermodynamic limit, and the dashed lines are extrapolations. 
}
\end{figure}

At zero temperature we extracted the critical point
from extrapolating the maxima of the von Neumann entropy, which in the thermodynamic limit should
diverge at the transition, for systems up to 22 ions and periodic 
boundary conditions. Experimentally, the location of the quantum phase transition can 
can be obtained for example from the Binder ratio~\cite{binder_critical_1981}.

On the ferromagnetic side there is a finite temperature transition for $\alpha<2$. For faster
decay there is a crossover~\cite{sachdev_quantum_2011}. At 
$\alpha=2$ and $h=0$, dashed lines, a Berezinskii-Kosterlitz-Thouless (BKT) transition occurs~\cite{thouless_long-range_1969,kosterlitz_phase_1976,cardy_one-dimensional_1981,bhattacharjee_properties_1981}, as the 
energy for a domain wall diverges logarithmically. The BKT transition also extends to finite transverse 
field $h$~\cite{dutta_phase_2001}. The symbols showing the finite temperature transition for $h=0$ ($h=J/2$) are 
(quantum) Monte-Carlo results~\cite{luijten_criticality_2001,sandvik_stochastic_2003}. We also calculate the location of the phase transition
using finite temperature Lanczos techniques~\cite{jaklic_lanczos_1994,aichhorn_low-temperature_2003}, which lead to an agreement to the quantum Monte-Carlo
within $10\%$. 

Critical exponents for spatial correlations 
extrapolated for systems with OBC up to $L=20$ are shown in \figc{fig:qc}{a}, symbols. For the short-ranged, 
transverse field Ising model it is well known that the Fisher exponent $\eta_x=1/4$, 
the critical exponent $z=1$ and thus we can deduce from \eq{eq:gzgx} $\eta_z=9/4$, which 
we recover well within our finite size extrapolations. For slowly decaying interactions
the system sizes are too small to recover the exact exponents, but a pronounced change
in the critical exponents when crossing over $\alpha=3$ can be observed. Exactly known values for the exponents are indicated 
by solid lines, whereas dashed lines are extrapolations between the Ising and the mean
field limit. 
Critical dynamic exponents $\eta_x z$ for $G^{xx,-}_{L/2,L/2}(t)$ are shown in \figc{fig:qc}{b},
as well as the dynamical exponent $z$ obtained from the ratio of the dynamic and the spatial
correlations, symbols. They agree reasonably with the exact results ($\alpha>3$ and $\alpha < 5/3$) 
even though the systems available to extrapolate are rather small.

\end{document}